\documentclass[12pt]{article}
\pdfoutput=1 
\usepackage{slashed}
\usepackage{color, verbatim}
\usepackage{latexsym}
\usepackage{amssymb}
\usepackage{amsmath}
\usepackage{graphicx}
\graphicspath{{figs/}}
\usepackage{arydshln}
\usepackage[dvipsnames]{xcolor}
\usepackage{adjustbox}
\usepackage{easybmat}
\usepackage{bbm}
\usepackage{cite}
\usepackage{slashed}
\usepackage{bm}
\usepackage{adjustbox}
\usepackage{booktabs}
\usepackage{multirow} 
\usepackage{rotating}
\usepackage{mathtools}
\usepackage[inline]{enumitem}
\usepackage{lscape}
\usepackage{booktabs}
\usepackage[normalem]{ulem}

\usepackage{hyperref}
\newcommand{\ii}{\ensuremath{\mathrm{i}}}
\renewcommand{\theequation}{\thesection.\arabic{equation}}
\makeatletter
\@addtoreset{equation}{section}
\g@addto@macro\bfseries{\boldmath}
\newcommand\Label[1]{&\refstepcounter{equation}(\theequation)\ltx@label{#1}&}
\makeatother
\allowdisplaybreaks

\begin{document}
%
\thispagestyle{empty}
\begin{flushright}
\end{flushright}
\vspace{0.8cm}

\begin{center}
{\Large\sc A Green's basis for the bosonic\\[0.5cm] SMEFT to dimension 8
  }
\vspace{0.8cm}

\textbf{
Mikael Chala$^{\,a}$, \'Alvaro D\'iaz-Carmona$^{\,a}$ and Guilherme Guedes$^{\,a,b}$}\\
\vspace{1.cm}
{\em {$^a$ CAFPE and Departamento de F\'isica Te\'orica y del Cosmos,
Universidad de Granada, Campus de Fuentenueva, E--18071 Granada, Spain}}\\[0.5cm]
{\em {$^b$ Laborat\'orio de Instrumenta\c cao e F\'isica Experimental de Part\'iculas, Departamento de
F\'isica da Universidade do Minho, Campus de Gualtar, 4710-057 Braga, Portugal}}\\[0.2cm]

\vspace{0.5cm}
\end{center}
\begin{abstract}
We present a basis of dimension-eight Green's functions involving Standard Model (SM) bosonic fields, consisting of $86$ new operators. Rather than using algebraic identities and integration by parts, we prove the independence of these interactions in momentum space, including a discussion on evanescent bosonic operators.  Our results pave the way for renormalising the SM effective field theory (SMEFT), as well as for performing matching of ultraviolet models onto the SMEFT, to higher order. To demonstrate the potential of our construction, we have implemented our basis in \texttt{matchmakereft} and used it to integrate out a heavy singlet scalar and a heavy quadruplet scalar up to one loop. We provide the corresponding dimension-eight Wilson coefficients. Likewise, we show how our results can be easily used to simplify cumbersome redundant Lagrangians arising, for example, from integrating out heavy fields using the path-integral approach to matching.
\end{abstract}

\newpage

\tableofcontents

\section{Introduction}
Effective field theories (EFTs) are indispensable tools for tackling quantum field theory (QFT) problems involving very different mass scales, particularly when employing mass-independent renormalisation schemes such as $\overline{\text{MS}}$.

The Lagrangian of an EFT at certain energy scale $\mu$ is organised as a power expansion in operators of increasing energy dimension (and therefore decreasing relevance). If some observable at a much smaller energy $\mu'\ll \mu$ is to be computed, the Lagrangian must be run down using the renormalisation group equations (RGEs). In doing so, one typically crosses energy thresholds corresponding to  particles of mass $M$. The latter do not decouple in mass-independent schemes; decoupling must be instead enforced by integrating out $M$. In practice, this is done by requiring that the Lagrangians with and without the particles of mass $M$ describe the same physics at $\mu=M$. This procedure is usually known as \textit{matching}.

It is well known that EFT operators related by (non-singular) field redefinitions are physically equivalent~\cite{Chisholm:1961tha,Kamefuchi:1961sb}. They give exactly the same $S$-matrix, which in turn requires all particles to be on the mass shell. However, when addressing the processes of running and matching, it is common practice to perform intermediate computations off-shell. In particular, for the matching this implies equating one-light-particle irreducible (off-shell) Green's functions above and below $\mu=M$, rather than $S$-matrix elements. 
There are a variety of reasons for proceeding this way.
\begin{enumerate}
 \item Within the traditional way of computing within QFT, namely \textit{\`a la Feynman}, $S$-matrix elements are obtained from connected and amputated diagrams. The amount of these scales much faster than one-particle-irreducible (1PI) diagrams with the number of external legs, thus turning the calculations computationally very challenging.
 \item In the path integral approach to matching~\cite{Gaillard:1985uh,Cheyette:1987qz,Henning:2014wua}, no off-shell Green's functions are computed. However, the resulting EFT involves, in general, operators related not only by field redefinitions, but even by algebraic identities, integration by parts, etc. The easiest way to simplify this EFT is by matching it \textit{off-shell} at tree level onto a set of independent Green's functions whose reduction to a physical basis must be known in advance. 
 \item While helicity-amplitude methods have been proved powerful to address the computation of some anomalous dimensions strictly on-shell~\cite{Caron-Huot:2016cwu,Bern:2019wie,Yang:2019vag,Baratella:2020lzz,Bern:2020ikv,EliasMiro:2020tdv,Baratella:2020dvw,Jiang:2020mhe,AccettulliHuber:2021uoa,Baratella:2021guc}, to the best of our knowledge they are not yet mature enough to be used in a number of cases. These include renormalisation triggered by operators with smaller number of legs than the resulting amplitude or mixing of amplitudes of different mass dimensions. 
\end{enumerate} 
The price to pay for working off-shell, though, is that operators related by field redefinitions must be kept in the action. Still, those related by algebraic identities (including Fierz relations) or integration by parts must be considered equivalent. A complete set of independent operators (up to field redefinitions) is called a \textit{Green's basis}.

The Standard Model EFT (SMEFT) is arguably the most important EFT, and likely the most reliable extension of the SM.~\footnote{This claim is supported by the lack of evidence of new particles between the electroweak scale $v\sim 246$ GeV and the TeV; as well as by the fact that the most tantalizing experimental anomalies, namely those in $B$ decays, point to scales of several TeVs.} A Green's basis for the SMEFT to dimension six was worked out in Ref.~\cite{Gherardi:2020det}. On top of the 84 real terms appearing in the physical basis~\cite{Grzadkowski:2010es} (for one fermion family), it involves 8 bosonic and 73 fermionic new operators, thus a total of 81 redundant interactions. (Green's bases are also known for the EFT of the SM with sterile neutrinos~\cite{Chala:2020vqp} to dimensions six, as well as for the EFT of the SM extended with an axion-like particle to dimension five~\cite{Chala:2020wvs}.)
 
However, there is by now convincing evidence that the dimension-six sector alone is not sufficient for making predictions within the SMEFT in a number of situations; see for example Refs.~\cite{Hays:2018zze,Chala:2018ari,Panico:2018hal,Ellis:2019zex,Alioli:2020kez,Gu:2020ldn,Ellis:2020ljj,Hays:2020scx,Corbett:2021eux,Ardu:2021koz} for details. Thus, in this article we elaborate a Green's basis for the SMEFT to dimension eight. Given the large number of operators (1649 on top of the 993 physical ones of Refs.~\cite{Murphy:2020rsh} and \cite{Li:2020gnx}) and the technical difficulties, we restrict this work to bosonic interactions. We find a total of 86 redundant operators that extend the 89 physical ones.

Let us notice that there are automatic tools to count independent off-shell operators, in particular \texttt{Basisgen}~\cite{Criado:2017khh} and  \texttt{Sym2Int}~\cite{Fonseca:2019yya}. However, to the best of our knowledge, none can yet build the operators explicitly, leave apart reducing them to the physical basis. Even checking whether certain operators within a set of interactions can be related by integration by parts, Fierz and Bianchi identities, etc. can become extremely cumbersone when the operators involve several fields and derivatives.

Thus, to simplify this task, we work in momentum space, where operator independence translate into the linear independence of 1PI tree-level amplitudes, conveniently written in terms of independent kinematic invariants. One of the most appreciated implications of this approach is that integration by parts reduces simply to momentum conservation. Likewise, Bianchi (and other) identities manifest simply and automatically as relations between different Feynman rules. 

This article is organised as follows. In Section~\ref{sec:theory} we introduce the SMEFT Lagrangian, including notation and sign conventions. In Section~\ref{sec:procedure} we discuss the approach to establishing the off-shell independence of interactions in momentum space. We also make emphasis on  subtleties arising in interactions involving pure four-dimensional objects such as the Levi-Civita symbol. In Section~\ref{sec:operators} we provide, class by class, the operators comprising the bosonic Green's basis. We also include the relevant information to cross-check their independence in momentum space. In Section~\ref{sec:onshell} we provide the explicit reduction of the redundant operators to the physical basis. Finally, with the aim of demonstrating the potential of intertwining our results with automatic tools, in Section~\ref{sec:applications} we integrate out a heavy scalar singlet as well as a heavy scalar quadruplet up to one loop to dimension eight using \texttt{matchmakereft}~\cite{Carmona:2021xtq}. We also show how complicated Lagrangians arising from matching using functional methods can be trivially reduced to a physical basis using our Green's set of operators together with standard tools such as \texttt{FeynRules}~\cite{Alloul:2013bka}. We conclude in Section~\ref{sec:conclusions}. Appendix~\ref{app:tables} includes tables with all operators for an easier reading. As auxiliary material, we provide the operators in a \texttt{FeynRules} model.

\section{Theory and conventions}
\label{sec:theory}
Given that we focus only on the bosonic sector of the SMEFT, the relevant renormalisable SM Lagrangian reads simply: 
\begin{align}\nonumber
 \mathcal{L}_\text{SM} = & -\frac{1}{4}G_{\mu\nu}^{A}G^{A\,\mu\nu} -\frac{1}{4}W_{\mu\nu}^{a}W^{a\,\mu\nu} -\frac{1}{4}B_{\mu\nu}B^{\mu\nu}\\\nonumber
 %
 %
& +\left(D_{\mu}\phi\right)^{\dagger}\left(D^{\mu}\phi\right)
-\mu^{2}|\phi|^{2}-\lambda|\phi|^{4}\,.
 %
\end{align}
%
We denote by $W, B$ and $G$ the electroweak gauge bosons and the gluon, respectively. We represent the Higgs doublet by $\phi =  (\varphi^+, \varphi^0)^T$, and $\tilde{\phi} = \ii\sigma_2\phi^*$ with $\sigma_I$ ($I=1,2,3$) being the Pauli matrices. Our convention for the covariant derivative is:
\begin{equation}
 D_\mu = \partial_\mu - \ii g_1 Y B_\mu -ig_2\frac{\sigma^I}{2} W_\mu^I -\ii g_3\frac{\lambda^A}{2} G_\mu^A\,,
\end{equation}
where $g_1, g_2$ and $g_3$ represent, respectively, the $U(1)_Y$, $SU(2)_L$ and $SU(3)_c$ gauge couplings, $Y$ stands for the hypercharge and $\lambda^A$ are the Gell-Mann matrices. Also, as customary, we define dual tensors as $\widetilde{X}_{\mu\nu}=\frac{1}{2} \epsilon_{\mu\nu\rho\lambda} X^{\rho\lambda}$, where our convention for the Levi-Civita symbol follows that of \texttt{FeynRules} and \texttt{FormCalc}~\cite{Hahn:1998yk}, $\epsilon_{0123}=+1$.

The SMEFT extendes $\mathcal{L}_\text{SM}$ with effective operators of energy dimension $d> 4$, suppressed by powers of the cutoff $\Lambda$, above which the SMEFT is no longer a valid EFT. Ignoring lepton number violation and operators of dimension higher than eight, the SMEFT Lagrangian reads:
\begin{equation}
 \mathcal{L}_\text{SMEFT} = \mathcal{L}_\text{SM} + \sum_i \frac{c_i^{(6)}}{\Lambda^2}\mathcal{O}_i^{(6)} + \sum_j \frac{c_j^{(8)}}{\Lambda^4}\mathcal{O}_j^{(8)}\,.
\end{equation}
Any set of physically independent dimension-six (bosonic) operators involve $15$ terms. A widely-used such set is the \textit{Warsaw} basis of Ref.~\cite{Grzadkowski:2010es}. Likewise, the bosonic sector of the SMEFT to dimension eight comprises $89$ independent physical couplings~\cite{Murphy:2020rsh,Li:2020gnx}.

In this work, though, we are interested in operators that are independent \textit{off-shell}, namely up to field redefinitions. A basis of operators of this kind is known to dimension six; see Ref.~\cite{Gherardi:2020det}.
The remaining of the paper is devoted to discussing the corresponding dimension-eight set, the way we have obtained it and some applications.
Hereafter, unless otherwise stated, \textit{independence} of operators will implicitly mean \textit{off-shell independence}.

\section{Off-shell independence in momentum space}
\label{sec:procedure}

Let $\lbrace\mathcal{O}_i\rbrace_{i=1...N}$ be $N$ (potentially dependent) operators, and let $\mathcal{A}(a\to b)$ be the 1PI amplitude for a given process $a\to b$. The contribution of $\mathcal{O}_i$ to the latter can be written in terms of \textit{independent} kinematics invariants $\lbrace \kappa_\alpha\rbrace_{\alpha\in I}$ where $I$ is a collection of indices. Explicitly, at tree level:
\begin{align}
 \mathcal{A}(a\to b) = c_i\sum_{\alpha\in I}  f^i_{\,\,\alpha}(\vec{g}) \kappa_\alpha\,,
\end{align}
where $f$ is simply a matrix that is a function of $\vec{g}=(g_1,g_2,g_3,\lambda)$, which collectively encodes the SM couplings. (No sum over $i$ is implied.)

If two operators $\mathcal{O}_i$ and $\mathcal{O}_j$, $i\neq j$, describe different off-shell physics, and precisely because the $\kappa$ are independent, then there must exist at least one process for which the corresponding $f_\alpha^i = (f_{\alpha_1}^i, f_{\alpha_2}^i, ...)$ and $f_\alpha^j = (f_{\alpha_1}^j, f_{\alpha_2}^j, ...)$ are non-collinear vectors.
In more generality, if there is one amplitude $\mathcal{A}$ such that the associated matrix $M$ with elements $(M)_{ij} = f^i_j$ has rank $N$, then the operators $\lbrace \mathcal{O}_i\rbrace_{i=1... N}$ are independent.
(Clearly, the opposite is not true.)

As a matter of example, let us consider the following dimension-eight six-Higgs operators:
\begin{align}
 \mathcal{O}_1 &= (\phi^\dagger\phi) D_\mu (\phi^\dagger\phi) D^\mu (\phi^\dagger\phi)\,,\\
 \mathcal{O}_2 &= (\phi^\dagger\phi)^2 (D^2\phi^\dagger\phi+\phi^\dagger D^2\phi)\,,\\
 \mathcal{O}_3 &= (\phi^\dagger\phi)^2 D_\mu\phi^\dagger D^\mu\phi\,.
\end{align}
They are all hermitian. The 1PI amplitude for $\varphi^0(p_1)\to\varphi^0 (p_2)\varphi^+(p_3) \varphi^-(p_4) \varphi^+(p_5) \varphi^- (p_6)$ reads:
\begin{align}\nonumber
 \mathcal{A} &= 2 i\, c_1 (2\kappa_{13}+2\kappa_{14}+2\kappa_{15}+2\kappa_{16}-2\kappa_{23}-2\kappa_{24}-2\kappa_{25}-2\kappa_{26}\\\nonumber&\,\,\,\,\,\,\,\,\,\,\,\,\,\,\,\,-\kappa_{34}-2\kappa_{35}-\kappa_{36}-\kappa_{45}-2\kappa_{46}-\kappa_{56})\\[0.1cm]\nonumber
 &- 4 i \,c_2 (\kappa_{11}+\kappa_{22}+\kappa_{33}+\kappa_{44}+\kappa_{55}+\kappa_{66})\\[0.1cm]
 &+ 2i\, c_3 (2\kappa_{12}-\kappa_{34}-\kappa_{36}-\kappa_{45}-\kappa_{56})\,,
\end{align}
where $\kappa_{ij} = p_i\cdot p_j$.

Clearly, the matrix $M$ associated to this process has, in appearance, rank 3. Indeed, for example the sub-matrix $\hat{M}$ associated to the invariants $\kappa_{11}$, $\kappa_{12}$, $\kappa_{13}$ looks like
\begin{equation}
 \hat{M} = \begin{bmatrix}0 & 0 & 4i\\ -4i & 0 & 0\\0 & 4i &0 \end{bmatrix}\Longrightarrow 3\geq\rm{Rank}(M)\geq \rm{Rank}(\hat{M}) = 3\,.
\end{equation}
However, the kinematic invariants that we have chosen are not independent, because by momentum conservation $p_1 = p_2+p_3+p_4+p_5+p_6$, and therefore $\kappa_{i1}$ can always be eliminated. Taking this into account, we get instead:
\begin{align}\nonumber
 \mathcal{A} &= 2i\,c_1 (2 \kappa_{33}+3 \kappa_{43}+2 \kappa_{44}+2 \kappa_{53} +3\kappa_{54}+2\kappa_{55}+3\kappa_{63}+2\kappa_{64}+3\kappa_{65}+2\kappa_{66} ) \\\nonumber
 &-8i\,c_2 (\kappa_{22}+\kappa_{32}+\kappa_{33}+\kappa_{42}+\kappa_{43}+\kappa_{44}+\kappa_{52}+\kappa_{53}+\kappa_{54}+\kappa_{55}+\kappa_{62}+\kappa_{63}\\\nonumber&\,\,\,\,\,\,\,\,\,\,\,\,\,+\kappa_{64}+\kappa_{65}+\kappa_{66})\\
 &+2i \, c_3 (2\kappa_{22}+2\kappa_{32}+2\kappa_{42}-\kappa_{43}+2\kappa_{52}-\kappa_{54}+2\kappa_{62}-\kappa_{63}-\kappa_{65})\,.
\end{align}
The corresponding matrix has only rank 2. This is clear from the fact that the first and third lines in the equation above add to minus half the second one. Or in other words, $\mathcal{O}_2 = -2(\mathcal{O}_1+\mathcal{O}_3)$. At the level of the Lagrangian, this results from the fact that the three operators are related by integration by parts up to a total derivative:
\begin{align}\nonumber
 D_\mu \bigg[(\phi^\dagger\phi)^2 D^\mu(\phi^\dagger\phi)\bigg] &= 2 (\phi^\dagger\phi) D_\mu(\phi^\dagger\phi) D^\mu (\phi^\dagger\phi)\\[-0.1cm]\nonumber
 &+ (\phi^\dagger\phi)^2 (D^2\phi^\dagger\phi+\phi^\dagger D^2\phi)\\[0.1cm]
 &+2 (\phi^\dagger\phi)^2 D_\mu\phi^\dagger D^\mu\phi\,.
\end{align}
Let us now study a slightly more elaborated example. Let us consider the following three operators:
\begin{align}
 \mathcal{O}_1 &= D_\mu (\phi^\dagger\phi) D^\nu B^{\mu\rho} B_{\nu\rho}\,,\\
 \mathcal{O}_2 &= (D^2\phi^\dagger\phi+\phi^\dagger D^2\phi) B^{\nu\rho} B_{\nu\rho}\,,\\
 \mathcal{O}_3 &= D_\mu\phi^\dagger D^\mu\phi B^{\nu\rho} B_{\nu\rho}\,.
\end{align}
The amplitude for $\varphi^0(p_1)\to \varphi^0(p_2) B(p_3) B(p_4)$ takes the form:
\begin{align}\nonumber
 \mathcal{A} &= -i c_1 (\kappa_{3334} + 2\kappa_{3434} + \kappa_{3444} -\kappa'_{4333}-2 \kappa'_{4334}-\kappa'_{4344})\\[0.1cm]\nonumber\,
 &+ 4i c_2 (2 \kappa_{2234} + 2\kappa_{2334} + 2\kappa_{2434} + \kappa_{3334} + 2\kappa_{3434}+\kappa_{3444}-2\kappa'_{4322}-2\kappa'_{4323}\\\nonumber
 &\,\,\,\,\,\,\,\,\,\,\,\,\,\,-2\kappa'_{4324}-\kappa'_{4333}-2\kappa-2\kappa'_{4334}-\kappa_{4344})\\[0.1cm]
 &-4i c_3 (\kappa_{2234}+\kappa_{2334}+\kappa_{2434}-\kappa'_{4322}-\kappa_{4323}-\kappa_{4324})\,;
\end{align}
where we have removed $p_1$ using momentum conservation, and the kinematic invariants are: $\kappa_{ijkl} = (\varepsilon_3\cdot\varepsilon_4)(p_i\cdot p_j)(p_k\cdot p_l)$ and $\kappa'_{ijkl} = (\varepsilon_3\cdot p_i)(\varepsilon_4\cdot p_j)(p_k\cdot p_l)$, with $\varepsilon$ representing a polarization vector. The rank of the corresponding matrix is only 2, so one of the operators is a linear combination of the other two. In fact, from the expression above it is obvious to check that $\mathcal{O}_1 = -\frac{1}{4}\mathcal{O}_2 -\frac{1}{2}\mathcal{O}_3$. This result reflects the following Lagrangian relation:
\begin{align}\nonumber
 \mathcal{O}_1 &= -D_\mu(\phi^\dagger\phi) D^\mu B^{\rho\nu} B_{\nu\rho} - D_\mu (\phi^\dagger\phi) D^\rho B^{\nu\mu} B_{\nu\rho}\\\nonumber
 &= -D_\mu(\phi^\dagger\phi) D^\mu B^{\rho\nu} B_{\nu\rho} - D_\mu (\phi^\dagger\phi) D^\rho B^{\mu\rho} B_{\nu\rho}\\\nonumber
 &= -D_\mu(\phi^\dagger\phi) D^\mu B^{\rho\nu} B_{\nu\rho} - \mathcal{O}_1\\\nonumber
 \Rightarrow \mathcal{O}_1 &= -\frac{1}{2} D_\mu(\phi^\dagger\phi) D^\mu B^{\rho\nu} B_{\nu\rho}\\\nonumber
 &= \frac{1}{2}D^2(\phi^\dagger\phi) B^{\rho\nu} B^{\nu\rho}-\frac{1}{2}D_\mu(\phi^\dagger\phi) B^{\rho\nu} D^\mu B_{\nu\rho}\,\\\nonumber
 &= \frac{1}{2}D^2(\phi^\dagger\phi) B^{\rho\nu} B_{\nu\rho} - \mathcal{O}_1\\\nonumber
 \Rightarrow \mathcal{O}_1 &=  \frac{1}{4} D^2(\phi^\dagger\phi) B^{\rho\nu} B_{\nu\rho}\,\\\nonumber
 &= -\frac{1}{4} (D^2\phi^\dagger\phi+ \phi^\dagger D^2\phi) B^{\nu\rho} B_{\nu\rho} - \frac{1}{4} (2 D_\mu\phi^\dagger D^\mu\phi) B^{\nu\rho} B_{\nu\rho}\\
 &= -\frac{1}{4} \mathcal{O}_2 - \frac{1}{2}\mathcal{O}_3\,.
\end{align}
In the first equality we have used the Bianchi identity $D_\mu B^{\nu\rho} + D_\nu B^{\rho\mu}+D_\rho B^{\mu\nu}=0$; in the second we have renamed indices as $\nu\leftrightarrow\rho$ in the last operator; in the fifth equality we have integrated by parts the derivative acting on $B^{\rho\nu}$; while in the penultimate equality we have simply expanded the derivative. In all steps, we have also taken into account that $B$ is anti-symmetric: $B^{\nu\rho} = -B^{\rho\nu}$.

These purposely-easy-to-follow example calculations might look trivial to address also from the Lagrangian (position space) point of view. However, things get significantly more complicated when not only a few but instead tens of operators are involved. More importantly, the examples above show how one can obtain the relations between \textit{dependent} operators. However, demonstrating that several operators are independent, at the Lagrangian level, implies proving that there is \textit{no single combination} of operations (integration by parts, Bianchi, etc.) that can relate any of them. In momentum space, though, one only needs to check, still, the rank of the corresponding matrix.

There is one last subtlety that one has to deal with in operators involving dual field strength tensors. The fully anti-symmetric Levi-Civita symbol $\epsilon^{\mu\nu\rho\lambda}$, which is a pure 4-dimensional object, fulfills in particular the Schouten identity:
\begin{equation}
 g_{\mu\nu}\epsilon_{\alpha\beta\gamma\delta} + g_{\mu\alpha}\epsilon_{\beta\gamma\delta\nu} + g_{\mu\beta}\epsilon_{\gamma\delta\nu\alpha} + g_{\mu\gamma}\epsilon_{\delta\nu\alpha\beta} + g_{\mu\delta}\epsilon_{\nu\alpha\beta\gamma} = 0\,.
\end{equation}
Related to this, the following relations involving an arbitrary rank-2 tensor $T$ and two field strength tensors $X$ and $F$ hold:
\begin{align}\label{eq:4dim1}
  T_{[\mu\nu]} X^\mu_{\,\,\rho} \widetilde{F}^{\nu\rho} &= T_{[\mu\nu]} \widetilde{X}^{\mu}_{\,\,\rho} F^{\nu\rho}\,,\\
 T_{\lbrace\mu\nu\rbrace} X^\mu_{\,\,\rho} \widetilde{F}^{\nu\rho} &= -T_{\lbrace\mu\nu\rbrace} \widetilde{X}^{\mu}_{\,\,\rho} F^{\nu\rho}+\frac{1}{2}T^{\mu}_\mu \widetilde{X}^{\nu\rho} F_{\nu\rho}\,,\\
 T_{\mu\nu} X^{\mu\rho}\widetilde{X}^\nu_{\,\,\rho} &= \frac{1}{4} T_\mu^\mu X^{\nu\rho}\widetilde{X}_{\nu\rho}\,,
\end{align}
where $[\mu\nu]$ and $\lbrace \mu\nu\rbrace$ denote, as usual, anti-symmetrisation and symmetrisation, respectively, with respect to the indices involved.
Certainly, these relations are much less immediate to take care of at the level of amplitudes than, for example, momentum conservation.
However, precisely because they are restrictions that hold only in $D=4$ space-time dimensions~\footnote{A fully anti-symmetric rank--4 tensor in $D>4$ space-time dimensions does not necessarily fulfill the identities above.}, they can be automatically enforced upon requiring all Lorentz vectors (momenta and polarizations) involved in the amplitude to be four-vectors. In practice, we simply demand that five or more Lorentz vectors can not be linearly independent. Let us show how this work with a simple example.

Let us consider the operators 
\begin{align}
 \mathcal{O}_1 &= i(D_\mu\phi^\dagger\sigma^I D_\nu\phi-D_\nu\phi^\dagger \sigma^I D_\mu\phi) B^{\mu}_{\,\,\rho} \widetilde{W}^{I\nu\rho} \,,\\
 \mathcal{O}_2 &= i(D_\mu\phi^\dagger\sigma^I D_\nu\phi-D_\nu\phi^\dagger \sigma^I D_\mu\phi) \widetilde{B}^{\mu}_{\,\,\rho} W^{I\nu\rho} \,.
\end{align}
They are obviously related by the identity in Eq.~\eqref{eq:4dim1}.

Now, let us compute the amplitude for $\varphi^0(p_1)\to \varphi^0(p_2) W^3(p_3) B(p_4)$. Already after imposing momentum conservation, we obtain:
\begin{align}\nonumber
 \mathcal{A} &= c_1 (-\kappa_{323443}-\kappa_{323444}+\kappa_{343424}+\kappa_{342334}+\kappa_{342344})\\
 &+ c_2 (-\kappa_{423433}-\kappa_{423434}-\kappa_{343423}+\kappa_{342433}+\kappa_{342434})\,;
\end{align}
where in this case, $\kappa_{ijklmn} = \epsilon(\varepsilon_i,p_j,p_k,p_l) (\varepsilon_m\cdot p_n)$ and $\kappa'_{ijklmn} = \epsilon(\varepsilon_i,\varepsilon_j,p_k,p_l) (p_m\cdot p_n)$. Naively, one would conclude that the operators are independent, because the rank of the matrix of the system is obviously 2.

Now, let us require that the three momenta (we have removed $p_1$) and the two polarization vectors can not be linearly independent in four space-time dimensions. In other words:
\begin{equation}
p_4 = a_1 \varepsilon_3+a_2\varepsilon_4 +a_3 p_2+a_4 p_3
\end{equation}
for some real numbers $a_i, i=1,...,4$. Upon using this constraint, the amplitude reads:
\begin{align}\nonumber
 \mathcal{A} &= (c_1+c_2) \bigg[a_3 \kappa_{342323} + a_3 (1+a_4) \kappa_{342323}+a_4(1+a_4) \kappa_{342333} \\
 &\,\,\,\,\,\,\,\,\,\,\,\,\,\,\,\,\,\,\,\,+a_1 a_3 \kappa_{342332} + a_1(1+2 a_4) \kappa_{342333} +a_2 a_4 \kappa_{342343} + a_1^2 \kappa_{342333} \bigg]\,.
\end{align}
From this expression, it is obvious that both operators are related (and identical). The difference between the two is actually an \textit{evanescent} term, which vanishes in $D=4$.

Equipped with this background, our strategy for finding minimal sets of Green's operators in a given class (defined by number of field strength tensors, derivatives and Higgs fields) consists simply in building all possible operators in the class, computing their contribution to a \textit{single} particular amplitude in which they are all independent~\footnote{For this we simply search for an amplitude whose associated rank equals the number of independent off-shell operators as provided by \texttt{Sim2Int} and \texttt{Basisgen}.}, and finally eliminating operators whose removal does not decrease the rank of the system.
From our point of view, automatising this procedure in momentum space is significantly simpler than in position space. 

\section{Explicit form of the operators}
\label{sec:operators}
Following \texttt{Basisgen} and \texttt{Sym2Int}, there are $2$ \textit{redundant} operators in the class $\phi^6 D^2$, $10$ in $\phi^4 D^4$, $1$ in $\phi^2 D^6$, $4$ in $X\phi^4 D^2$, $44$ in $X^2\phi^2 D^2$, $6$ in $X\phi^2 D^4$, $16$ in $X^3 D^2$ and $3$ in the class $X^2 D^4$. Those in the first four classes have been already presented in Ref.~\cite{Chala:2021pll}. Hence, we focus here on the remaining operators.

For clarity, in the lists of interactions below, we include also the physical operators (in the classes in which they exist) as given in Ref.~\cite{Murphy:2020rsh}, and with the same names.

\subsection{Operators in the class $X \phi^2 D^4$}
There are 3 real terms for $X=B$, and 3 more for $X=W$.
In the first case, it suffices to compute the amplitude for the process $\varphi^0(p_1)\to\varphi^0(p_2) B(p_3)$, while in the second case only $\varphi^0(p_1)\to\varphi^0(p_2) W^+(p_2)$ is needed.


\subsubsection{$X=B$}
%
%
\begin{align}
 \mathcal{O}_{B\phi^2 D^4}^{(1)} &= i (D_\nu\phi^\dagger D^2\phi-D^2\phi^\dagger D_\nu\phi) D_\mu B^{\mu\nu}\,,\\
\mathcal{O}_{B\phi^2 D^4}^{(2)} &= (D_\nu\phi^\dagger D^2\phi+D^2\phi^\dagger D_\nu\phi) D_\mu B^{\mu\nu}\,,\\
\mathcal{O}_{B\phi^2 D^4}^{(3)} &= i (D_\rho D_\nu\phi^\dagger D^\rho\phi-D^\rho\phi^\dagger D_\rho D_\nu\phi) D_\mu B^{\mu\nu}\,.
\end{align}
%
%
%
\subsubsection{$X=W$}
%
\begin{align}
 \mathcal{O}_{W\phi^2 D^4}^{(1)} &= i (D_\nu\phi^\dagger \sigma^I D^2\phi-D^2\phi^\dagger\sigma^I D_\nu\phi) D_\mu W^{I\mu\nu}\,,\\
  \mathcal{O}_{W\phi^2 D^4}^{(2)} &= (D_\nu\phi^\dagger \sigma^I D^2\phi+D^2\phi^\dagger\sigma^I D_\nu\phi) D_\mu W^{I\mu\nu}\,,\\
 \mathcal{O}_{B\phi^2 D^4}^{(3)} &= i (D_\rho D_\nu\phi^\dagger\sigma^I D^\rho\phi-D^\rho\phi^\dagger \sigma^I D_\rho D_\nu\phi) D_\mu W^{I\mu\nu}\,.
\end{align}
%
%

\subsection{Operators in the class $X^2 \phi^2 D^2$}
There are 12 independent operators for $X^2=B^2$, $19$ for $X^2=W^2$ and also $19$ for $X^2=WB$.
One can check the independence of the operators below by evaluating the amplitudes $\varphi^0(p_1)\to\varphi^0(p_2)B(p_3)B(p_4)$,  $\varphi^0(p_1)\to\varphi^0(p_2)W^+(p_3)W^-(p_4)$ and  $\varphi^0(p_1)\to\varphi^+(p_2)W^+(p_3)B(p_4)$, respectively.

\subsubsection{$X^2 = B^2$}
\begin{align}
 \mathcal{O}_{B^2\phi^2 D^2}^{(1)} &= (D^\mu\phi^\dagger D_\nu\phi) B_{\mu\rho} B^{\nu\rho}\,,\\
 \mathcal{O}_{B^2\phi^2 D^2}^{(2)} &= (D^\mu\phi^\dagger D_\mu\phi) B_{\nu\rho} B^{\nu\rho}\,,\\
 \mathcal{O}_{B^2\phi^2 D^2}^{(3)} &= (D^\mu\phi^\dagger D_\mu\phi) B_{\nu\rho} \widetilde{B}^{\nu\rho}\,,\\
 \mathcal{O}_{B^2\phi^2 D^2}^{(4)} &= (D_\mu\phi^\dagger \phi+\phi^\dagger D_\mu\phi) D_\nu B^{\mu\rho} B^\nu_{\,\,\rho}\,,\\
 \mathcal{O}_{B^2\phi^2 D^2}^{(5)} &= i (\phi^\dagger D_\mu D_\nu\phi-D_\mu D_\nu \phi^\dagger \phi) B^{\mu\rho} B^{\nu}_{\,\,\rho}\,,\\
\mathcal{O}_{B^2\phi^2 D^2}^{(6)} &= \phi^\dagger\phi D_\mu D_\nu B^{\mu\rho} B^{\nu}_{\,\,\rho}\,,\\
\mathcal{O}_{B^2\phi^2 D^2}^{(7)} &= i (\phi^\dagger D_\nu\phi-D_\nu\phi^\dagger\phi) D_\mu B^{\mu\rho} B^{\nu}_{\,\,\rho}\,,\\
\mathcal{O}_{B^2\phi^2 D^2}^{(8)} &= (\phi^\dagger D_\nu\phi+D_\nu\phi^\dagger\phi) D_\mu B^{\mu\rho} B^\nu_{\,\,\rho}\,,\\
\mathcal{O}_{B^2\phi^2 D^2}^{(9)} &= (\phi^\dagger D^2\phi+D^2\phi^\dagger\phi) B^{\nu\rho} \widetilde{B}_{\nu\rho}\,,\\
\mathcal{O}_{B^2\phi^2 D^2}^{(10)} &= i (\phi^\dagger D^2\phi-D^2\phi^\dagger\phi) B^{\nu\rho} \widetilde{B}_{\nu\rho}\,\\
\mathcal{O}_{B^2\phi^2 D^2}^{(11)} &= (\phi^\dagger D_\nu\phi+D_\nu\phi^\dagger\phi) D_\mu B^{\mu\rho} \widetilde{B}^{\nu}_{\,\,\rho}\,\\
\mathcal{O}_{B^2\phi^2 D^2}^{(12)} &= i(\phi^\dagger D_\nu\phi - D_\nu\phi^\dagger\phi) D_\mu B^{\mu\rho} \widetilde{B}^{\nu}_{\,\,\rho}\,.
\end{align}

\subsubsection{$X^2 = W^2$}
\begin{align}
 \mathcal{O}_{W^2\phi^2 D^2}^{(1)} &= (D^\mu\phi^\dagger D^\nu\phi) W_{\mu\rho}^I W_\nu^{I\rho}\,,\\
 \mathcal{O}_{W^2\phi^2 D^2}^{(2)} &= (D^\mu\phi^\dagger D_\mu\phi) W_{\nu\rho}^I W^{I\nu\rho}\,,\\
 \mathcal{O}_{W^2\phi^2 D^2}^{(3)} &= (D^\mu\phi^\dagger D_\mu\phi) W_{\nu\rho}^I \widetilde{W}^{I\nu\rho}\,,\\
 \mathcal{O}_{W^2\phi^2 D^2}^{(4)} &= i\epsilon^{IJK}(D^\mu\phi^\dagger\sigma^I D^\nu\phi) W_{\mu\rho}^J W_\nu^{K\rho}\,,\\
 \mathcal{O}_{W^2\phi^2 D^2}^{(5)} &= \epsilon^{IJK}(D^\mu\phi^\dagger\sigma^I D^\nu\phi) (W_{\mu\rho}^J \widetilde{W}_\nu^{K\rho}-\widetilde{W}^J_{\mu\rho} W_\nu^{K\rho})\,,\\
 \mathcal{O}_{W^2\phi^2 D^2}^{(6)} &= i\epsilon^{IJK}(D^\mu\phi^\dagger\sigma^I D^\nu\phi) (W_{\mu\rho}^J \widetilde{W}_\nu^{K\rho}+\widetilde{W}^J_{\mu\rho} W_\nu^{K\rho})\,,\\
 \mathcal{O}_{W^2\phi^2 D^2}^{(7)} &= i\epsilon^{IJK}(\phi^\dagger \sigma^I D^\nu\phi-D^\nu\phi^\dagger \sigma^I \phi) D_\mu W^{J\mu\rho}\widetilde{W}^{K}_{\nu\rho}\,,\\
 \mathcal{O}_{W^2\phi^2 D^2}^{(8)} &= \epsilon^{IJK} \phi^\dagger\sigma^I\phi D_\nu D_\mu W^{J\mu\rho}\widetilde{W}^{K\nu}_{\,\,\,\,\rho}\,,\\
 \mathcal{O}_{W^2\phi^2 D^2}^{(9)} &= i (\phi^\dagger D_\nu\phi-D_\nu\phi^\dagger \phi) D_\mu W^{I\mu\rho} \widetilde{W}^{I\nu}_{\,\,\,\,\rho}\,,\\
 \mathcal{O}_{W^2\phi^2 D^2}^{(10)} &= (\phi^\dagger D_\nu\phi+D_\nu\phi^\dagger \phi) D_\mu W^{I\mu\rho} \widetilde{W}^{I\nu}_{\,\,\,\,\rho}\,,\\
 \mathcal{O}_{W^2\phi^2 D^2}^{(11)} &= (\phi^\dagger D_\nu\phi+D_\nu\phi^\dagger \phi) D_\mu W^{I\mu\rho} W^{I\nu}_{\,\,\,\,\rho}\,,\\
 \mathcal{O}_{W^2\phi^2 D^2}^{(12)} &= i(\phi^\dagger D_\nu\phi-D_\nu\phi^\dagger \phi) D_\mu W^{I\mu\rho} W^{I\nu}_{\,\,\,\,\rho}\,,\\
 \mathcal{O}_{W^2\phi^2 D^2}^{(13)} &= \phi^\dagger\phi D_\mu W^{I\mu\rho} D_\nu W^{I\nu}_{\,\,\,\,\rho}\,,\\
 \mathcal{O}_{W^2\phi^2 D^2}^{(14)} &= (D_\mu\phi^\dagger\phi+\phi^\dagger D_\mu\phi) W^{I\nu\rho} D^\mu W^{I}_{\nu\rho}\,,\\
 \mathcal{O}_{W^2\phi^2 D^2}^{(15)} &= i(D_\mu\phi^\dagger\phi-\phi^\dagger D_\mu\phi) W^{I\nu\rho} D^\mu W^{I}_{\nu\rho}\,,\\
 \mathcal{O}_{W^2\phi^2 D^2}^{(16)} &= (D_\mu\phi^\dagger\phi+\phi^\dagger D_\mu\phi) D^\mu W^{I\nu\rho}\widetilde{W}^{I}_{\nu\rho}\,,\\
 \mathcal{O}_{W^2\phi^2 D^2}^{(17)} &= i(D_\mu\phi^\dagger\phi-\phi^\dagger D_\mu\phi) D^\mu W^{I\nu\rho}\widetilde{W}^{I}_{\nu\rho}\,,
 \\
 \mathcal{O}_{W^2\phi^2 D^2}^{(18)} &= \epsilon^{IJK}(\phi^\dagger \sigma^I D^\nu\phi+D^\nu\phi^\dagger \sigma^I \phi) D_\mu W^{J\mu\rho}W^{K}_{\nu\rho}\,,\\
 \mathcal{O}_{W^2\phi^2 D^2}^{(19)} &= i\epsilon^{IJK}(\phi^\dagger \sigma^I D^\nu\phi-D^\nu\phi^\dagger \sigma^I \phi) D_\mu W^{J\mu\rho}W^{K}_{\nu\rho}\,.\\
\end{align}

\subsubsection{$X^2 = WB$}
\begin{align}
 \mathcal{O}_{WB\phi^2D^2}^{(1)} &= (D^\mu\phi^\dagger\sigma^I D_\mu\phi) B_{\nu\rho}W^{I\nu\rho}\,,\\
 \mathcal{O}_{WB\phi^2D^2}^{(2)} &= (D^\mu\phi^\dagger\sigma^I D_\mu\phi) B_{\nu\rho}\widetilde{W}^{I\nu\rho}\,,\\
 \mathcal{O}_{WB\phi^2D^2}^{(3)} &= i(D^\mu\phi^\dagger\sigma^I D^\nu\phi) (B_{\mu\rho}W^{I\,\rho}_{\nu}-B_{\nu\rho} W^{I\,\rho}_\mu)\,,\\
 \mathcal{O}_{WB\phi^2D^2}^{(4)} &= (D^\mu\phi^\dagger\sigma^I D^\nu\phi) (B_{\mu\rho}W^{I\,\rho}_{\nu}+B_{\nu\rho} W^{I\,\rho}_\mu)\,,\\
 \mathcal{O}_{WB\phi^2D^2}^{(5)} &= i(D^\mu\phi^\dagger\sigma^I D^\nu\phi) (B_{\mu\rho}\widetilde{W}^{I\,\rho}_{\nu}-B_{\nu\rho} \widetilde{W}^{I\,\rho}_\mu)\,,\\
 \mathcal{O}_{WB\phi^2D^2}^{(6)} &=  (D^\mu\phi^\dagger\sigma^I D^\nu\phi) (B_{\mu\rho}\widetilde{W}^{I\,\rho}_{\nu}+B_{\nu\rho} \widetilde{W}^{I\,\rho}_\mu)\,,\\
 \mathcal{O}_{WB\phi^2D^2}^{(7)} &= i(\phi^\dagger\sigma^I D^\mu\phi-D^\mu\phi^\dagger\sigma^I\phi) D_\mu B^{\nu\rho} W^{I}_{\nu\rho}\,,\\
 \mathcal{O}_{WB\phi^2D^2}^{(8)} &= (\phi^\dagger\sigma^I D^\nu\phi+D^\nu\phi^\dagger\sigma^I\phi) D_\mu B^{\mu\rho} W^{I}_{\nu\rho}\,,\\
 \mathcal{O}_{WB\phi^2D^2}^{(9)} &= i(\phi^\dagger\sigma^I D^\nu\phi-D^\nu\phi^\dagger\sigma^I\phi) D_\mu B^{\mu\rho} W^{I}_{\nu\rho}\,,\\
 \mathcal{O}_{WB\phi^2D^2}^{(10)} &= (\phi^\dagger\sigma^I\phi) D^\mu B_{\mu\rho} D_\nu W^{I\nu\rho} \,,\\
 \mathcal{O}_{WB\phi^2D^2}^{(11)} &= (D_\nu\phi^\dagger\sigma^I\phi+\phi^\dagger\sigma^I D_\nu\phi) B_{\mu\rho} D^\mu W^{I\nu\rho}\,,\\
 \mathcal{O}_{WB\phi^2D^2}^{(12)} &= i(D_\nu\phi^\dagger\sigma^I\phi-\phi^\dagger\sigma^I D_\nu\phi) B_{\mu\rho} D^\mu W^{I\nu\rho}\,,\\
 \mathcal{O}_{WB\phi^2D^2}^{(13)} &= (\phi^\dagger\sigma^I\phi) B_{\mu\rho}  D_\nu D^\mu W^{I\nu\rho} \,,\\
 \mathcal{O}_{WB\phi^2D^2}^{(14)} &= i(D_\nu\phi^\dagger\sigma^I\phi-\phi^\dagger\sigma^I D_\nu\phi) D^\mu B_{\mu\rho} \widetilde{W}^{I\nu\rho} \,,\\
 \mathcal{O}_{WB\phi^2D^2}^{(15)} &= i (\phi^\dagger\sigma^I D_\mu\phi-D_\mu\phi^\dagger\sigma^I\phi) D^\mu B_{\nu\rho} \widetilde{W}^{I\nu\rho}\,,\\
 \mathcal{O}_{WB\phi^2D^2}^{(16)} &= (\phi^\dagger\sigma^I\phi) (D^2 B^{\nu\rho}) \widetilde{W}^{I}_{\nu\rho}\,,\\
 \mathcal{O}_{WB\phi^2D^2}^{(17)} &= (\phi^\dagger\sigma^I\phi) (D^\rho D_\mu W^{I\mu\nu}) \widetilde{B}_{\nu\rho}\,,\\
 \mathcal{O}_{WB\phi^2D^2}^{(18)} &= i(D^\nu\phi^\dagger\sigma^I\phi-\phi^\dagger\sigma^I D^\nu\phi) \widetilde{B}^{\mu\rho} D_\mu W^I_{\nu\rho}\,,\\
 \mathcal{O}_{WB\phi^2D^2}^{(19)} &= (D^\nu\phi^\dagger\sigma^I\phi+\phi^\dagger\sigma^I D^\nu\phi) \widetilde{B}^{\mu\rho} D_\mu W^I_{\nu\rho}\,.
\end{align}

\subsubsection{$X^2 = G^2$}
\begin{align}
 \mathcal{O}_{G^2\phi^2 D^2}^{(1)} &= (D^\mu\phi^\dagger D_\nu\phi) G^A_{\mu\rho} G^{A\nu\rho}\,,\\
 \mathcal{O}_{G^2\phi^2 D^2}^{(2)} &= (D^\mu\phi^\dagger D_\mu\phi) G^A_{\nu\rho} G^{A\nu\rho}\,,\\
 \mathcal{O}_{G^2\phi^2 D^2}^{(3)} &= (D^\mu\phi^\dagger D_\mu\phi) G^A_{\nu\rho} \widetilde{G}^{A\nu\rho}\,,\\
 \mathcal{O}_{G^2\phi^2 D^2}^{(4)} &= (D_\mu\phi^\dagger \phi+\phi^\dagger D_\mu\phi) D_\nu G^{A\mu\rho} G^{A\nu}_{\,\,\rho}\,,\\
 \mathcal{O}_{G^2\phi^2 D^2}^{(5)} &= i (\phi^\dagger D_\mu D_\nu\phi-D_\mu D_\nu \phi^\dagger \phi) G^{A\mu\rho} G^{A\nu}_{\,\,\rho}\,,\\
\mathcal{O}_{G^2\phi^2 D^2}^{(6)} &= \phi^\dagger\phi D_\mu D_\nu G^{A\mu\rho} G^{A\nu}_{\,\,\rho}\,,\\
\mathcal{O}_{G^2\phi^2 D^2}^{(7)} &= i (\phi^\dagger D_\nu\phi-D_\nu\phi^\dagger\phi) D_\mu G^{A\mu\rho} G^{A\nu}_{\,\,\rho}\,,\\
\mathcal{O}_{G^2\phi^2 D^2}^{(8)} &= (\phi^\dagger D_\nu\phi+D_\nu\phi^\dagger\phi) D_\mu G^{A\mu\rho} G^{A\nu}_{\,\,\rho}\,,\\
\mathcal{O}_{G^2\phi^2 D^2}^{(9)} &= (\phi^\dagger D^2\phi+D^2\phi^\dagger\phi) G^{A\nu\rho} \widetilde{G}^{A\nu}_{\,\,\rho}\,,\\
\mathcal{O}_{G^2\phi^2 D^2}^{(10)} &= i (\phi^\dagger D^2\phi-D^2\phi^\dagger\phi) G^{A\nu\rho} \widetilde{G}^A_{\nu\rho}\,\\
\mathcal{O}_{G^2\phi^2 D^2}^{(11)} &= (\phi^\dagger D_\nu\phi+D_\nu\phi^\dagger\phi) D_\mu G^{A\mu\rho} \widetilde{G}^{A\nu}_{\,\,\rho}\,\\
\mathcal{O}_{G^2\phi^2 D^2}^{(12)} &= i(\phi^\dagger D_\nu\phi - D_\nu\phi^\dagger\phi) D_\mu G^{A\mu\rho} \widetilde{G}^{A\nu}_{\,\,\rho}\,.
\end{align}

\subsection{Operators in the class $X^3 D^2$}
In this case, there are 4 operators for each of the combinations $X^3=W^2B$, $X^3=G^2B$, $X^3=W^3$ and $X^3=G^3$. The (CP-conserving) $W^3$ and $G^3$ operators were previously presented in Ref.~\cite{Quevillon:2018mfl}. For the test, again, only one amplitude is needed for each combination to manifest their independence. For example: $B(p_1)\to W^+(p_2) W^-(p_3)$ and $B(p_1)\to G(p_2) G(p_3)$.
\subsubsection{$X^3=W^2 B$}
\begin{align}
 \mathcal{O}_{W^2B D^2}^{(1)} &= B_{\mu\nu} D_\rho W^{I\mu\nu} D_\sigma W^{I\rho\sigma}\,,\\
 \mathcal{O}_{W^2B D^2}^{(2)} &= B_{\mu\nu} (D^2 W^{I\mu\rho}) W^{I\nu}_{\,\,\,\,\,\,\rho}\,,\\
 \mathcal{O}_{W^2B D^2}^{(3)} &= \widetilde{B}_{\mu\nu} D_\rho W^{I\mu\nu} D_\sigma W^{I\rho\sigma}\,,\\
 \mathcal{O}_{W^2B D^2}^{(4)} &= \widetilde{B}_{\mu\nu} (D^2 W^{I\mu\rho}) W^{I\nu}_{\,\,\,\,\,\,\rho}\,.
\end{align}

\subsubsection{$X^3=G^2 B$}
\begin{align}
 \mathcal{O}_{G^2B D^2}^{(1)} &= B_{\mu\nu} D_\rho G^{A\mu\nu} D_\sigma G^{A\rho\sigma}\,,\\
 \mathcal{O}_{G^2B D^2}^{(2)} &= B_{\mu\nu} (D^2 G^{A\mu\rho}) G^{A\nu}_{\,\,\,\,\,\,\rho}\,,\\
 \mathcal{O}_{G^2B D^2}^{(3)} &= \widetilde{B}_{\mu\nu} D_\rho G^{A\mu\nu} D_\sigma G^{A\rho\sigma}\,,\\
 \mathcal{O}_{G^2B D^2}^{(4)} &= \widetilde{B}_{\mu\nu} (D^2 G^{A\mu\rho}) G^{A\nu}_{\,\,\,\,\,\,\rho}\,.
\end{align}

\subsubsection{$X^3 = W^3$}
\begin{align}
 \mathcal{O}_{W^3 D^2}^{(1)} &= \epsilon^{IJK} W^I_{\mu\nu} D_\rho W^{J\mu\nu} D_\sigma W^{K\rho\sigma}\,,\\
\mathcal{O}_{W^3 D^2}^{(2)} &= \epsilon^{IJK} W^I_{\mu\nu} D_\rho W^{J\rho\mu} D_\sigma W^{K\sigma\nu}\,,\\
 \mathcal{O}_{W^3 D^2}^{(3)} &= \epsilon^{IJK}\widetilde{W}^I_{\mu\nu} D_\rho W^{J\mu\nu} D_\sigma W^{K\rho\sigma}\,,\\
\mathcal{O}_{W^3 D^2}^{(4)} &= \epsilon^{IJK} \widetilde{W}^I_{\mu\nu} D_\rho W^{J\rho\mu} D_\sigma W^{K\sigma\nu}\,,\\
\end{align}

\subsubsection{$X^3 = G^3$}
\begin{align}
 \mathcal{O}_{G^3 D^2}^{(1)} &= f^{ABC} G^A_{\mu\nu} D_\rho G^{B\mu\nu} D_\sigma G^{C\rho\sigma}\,,\\
\mathcal{O}_{G^3 D^2}^{(2)} &= f^{ABC} G^A_{\mu\nu} D_\rho G^{B\rho\mu} D_\sigma G^{C\sigma\nu} \;, \\
 \mathcal{O}_{G^3 D^2}^{(3)} &= f^{ABC}\widetilde{G}^A_{\mu\nu} D_\rho G^{B\mu\nu} D_\sigma G^{C\rho\sigma}\,,\\
\mathcal{O}_{G^3 D^2}^{(4)} &= f^{ABC} \widetilde{G}^A_{\mu\nu} D_\rho G^{B\rho\mu} D_\sigma G^{C\sigma\nu} \;, 
\end{align}

\subsection{Operators in the class $X^2 D^4$}
In this class, there is only 1 operator per category, $X=B,W,G$. So the independence of operators is obvious.

\subsubsection{$X=B$}
\begin{align}
 \mathcal{O}_{B^2 D^4} &= (D_\sigma D_\mu B^{\mu\nu}) (D^\sigma D^\rho B_{\rho\nu})\,.
\end{align}

\subsubsection{$X=W$}
\begin{align}
 \mathcal{O}_{W^2 D^4} &= (D_\sigma D_\mu W^{I\mu\nu}) (D^\sigma D^\rho W^I_{\rho\nu})\,.
\end{align}

\subsubsection{$X=G$}
\begin{align}
 \mathcal{O}_{G^2 D^4} &= (D_\sigma D_\mu G^{A\mu\nu}) (D^\sigma D^\rho G^A_{\rho\nu})\,.
\end{align}

\section{On-shell relations}
\label{sec:onshell}
Tables~\ref{tab:dim8ops1}, \ref{tab:dim8ops2}, \ref{tab:dim8ops3} and \ref{tab:dim8ops4} show all bosonic operators comprising the Green's basis for the dimension-eight SMEFT. Those in gray are redundant on-shell. They can be reduced to the physical basis of Ref.~\cite{Murphy:2020rsh} by using the SM equations of motion. Ignoring fermions, the latter read simply:
\begin{align}
 D^2\phi^i &= \mu^2\phi^i -2\lambda(\phi^\dagger\phi)\phi^i\,,\\
 \partial^\nu B_{\mu\nu} &=\frac{g_1}{2}(\phi^\dagger iD_\mu\phi-D_\mu\phi^\dagger i \phi)\,,\\
 D^\nu W^{I}_{\mu\nu} &= \frac{g_2}{2}(\phi^\dagger i D_\mu\sigma^I\phi-\phi^\dagger i\sigma^I D_\mu\phi)\,.
\end{align}
The following relations are also useful:
\begin{align}
 [D_\mu, D_\nu]\phi &= -i\frac{g_1}{2} B_{\mu\nu}\phi -i \frac{g_2}{2}\sigma^I W^{I}_{\mu\nu}\phi\,,\\
 [D_\mu, D_\nu] W^{I\rho\lambda} &= g_2 \epsilon^{IJK} W^{J}_{\mu \nu} W^{K\rho\lambda}\,,\\
 [D_\mu, D_\nu] G^{A\rho\lambda} &= g_3 f^{ABC} G^{B}_{\mu \nu} G^{C\rho\lambda}\,. 
\end{align}
The operators $\mathcal{O}_{\phi^4}^{(5)}$,$\mathcal{O}_{\phi^4}^{(7)}$,$\mathcal{O}_{\phi^4}^{(9)}$,$\mathcal{O}_{\phi^4}^{(13)}$, $\mathcal{O}_{\phi^6}^{(4)}$, $\mathcal{O}_{B\phi^2D^4}^{(2)}$, $\mathcal{O}_{W\phi^2D^4}^{(2)}$, $\mathcal{O}_{W\phi^2D^4}^{(5)}$, $\mathcal{O}_{W\phi^4D^2}^{(5)}$, $\mathcal{O}_{B^2\phi^2D^2}^{(5)}$, $\mathcal{O}_{B^2\phi^2D^2}^{(7)}$, $\mathcal{O}_{B^2\phi^2D^2}^{(10)}$, $\mathcal{O}_{B^2\phi^2 D^2}^{(12)}$, $\mathcal{O}_{W^2\phi^2D^2}^{(15)}$, $\mathcal{O}_{W^2\phi^2D^2}^{(17)}$, $\mathcal{O}_{G^2\phi^2D^2}^{(5)}$, $\mathcal{O}_{G^2\phi^2D^2}^{(7)}$, $\mathcal{O}_{G^2\phi^2D^2}^{(8)}$, $\mathcal{O}_{G^2\phi^2D^2}^{(10)}$, $\mathcal{O}_{G^2\phi^2 D^2}^{(11)}$, $\mathcal{O}_{G^2\phi^2 D^2}^{(12)}$, $\mathcal{O}_{G^2BD^2}^{(1)}$, $\mathcal{O}_{G^2BD^2}^{(2)}$, $\mathcal{O}_{G^2BD^2}^{(3)}$, $\mathcal{O}_{G^2BD^2}^{(4)}$, $\mathcal{O}_{G^3D^2}^{(1)}$, $\mathcal{O}_{G^3D^2}^{(2)}$, $\mathcal{O}_{G^3D^2}^{(3)}$, $\mathcal{O}_{G^3D^2}^{(4)}$ and $\mathcal{O}_{G^2D^4}$ vanish on-shell (up to fermionic interactions). The rest shift the Wilson coefficients of physical operators, which finally read as follows:
\begin{align}\label{eq:reductionI}
 c_{\phi^8} &\rightarrow c_{\phi^8} -\frac{1}{2} c_{B^2D^4} g_1^2 g_2^2 \lambda
   +\frac{1}{2} c_{B^2\phi^2D^2}^{(8)} g_1^2 \lambda +2
   c_{B\phi^2D^4}^{(1)} g_1 \lambda ^2+\frac{1}{4}
   c_{B\phi^2D^4}^{(3)} g_1 g_2^2 \lambda \, \nonumber \\
& -2
   c_{B\phi^2D^4}^{(3)} g_1 \lambda^2+c_{B\phi^4D^2}^{(3)} g_1
   \lambda +4 c_{\phi^4}^{(10)} \lambda^2+4 c_{\phi^4}^{(11)} \lambda^2
-4 c_{\phi^4}^{(12)} \lambda^2+8 c_{\phi^4}^{(8)} \lambda^2 \, \nonumber \\
&
-c_{W^2D^4} g_1^2 g_2^2 \lambda
   -\frac{1}{2} c_{W^2D^4} g_2^4 \lambda +\frac{1}{2}
   c_{W^2\phi^2D^2}^{(11)} g_2^2 \lambda -\frac{1}{2}
   c_{W^2\phi^2D^2}^{(13)} g_2^2 \lambda -c_{W^2\phi^2D^2}^{(19)}
   g_2^2 \lambda \,\nonumber \\
& -c_{W^3D^2}^{(1)} g_2^3 \lambda
   +\frac{1}{2} c_{W^3D^2}^{(2)} g_2^3 \lambda -\frac{1}{2}
   c_{WB\phi^2D^2}^{(10)} g_1 g_2 \lambda
   +c_{WB\phi^2D^2}^{(13)} g_1g_2 \lambda -\frac{1}{2}
   c_{WB\phi^2D^2}^{(8)} g_1 g_2 \lambda \, \nonumber \\
& +2
   c_{W\phi^2D^4}^{(1)} g_2 \lambda ^2+\frac{1}{2}
   c_{W\phi^2D^4}^{(3)} g_1^2 g_2 \lambda -2
   c_{W\phi^2D^4}^{(3)} g_2 \lambda ^2+c_{W\phi^4D^2}^{(6)} g_2
   \lambda +\frac{c_{W\phi^4D^2}^{(7)} g_2 \lambda }{2}\, \nonumber \\
&-4 c_{\phi^6}^{(3)} \lambda -c_{\phi^2} \left( g_1^2 \lambda^2 +
 g_2^2 \lambda^2 + 32 \lambda^3\right)
 \,,\\
 c_{\phi^6}^{(1)} &\rightarrow c_{\phi^6}^{(1)} + \frac{3}{2} c_{B^2D^4} g_1^2 g_2^2-\frac{3
   c_{B^2\phi^2D^2}^{(8)}  g_1^2}{4}-3 c_{B\phi^2D^4}^{(1)}  g_1
   \lambda -\frac{3}{4} c_{B\phi^2D^4}^{(3)}  g_1  g_2^2+3
   c_{B\phi^2D^4}^{(3)}  g_1 \lambda \,\nonumber \\
& -\frac{3 c_{B\phi^4D^2}^{(3)}
    g_1}{2}+\frac{3}{2} c_{\phi^2}  g_1^2 \lambda
   +\frac{5}{2} c_{\phi^2}  g_2^2 \lambda +8 c_{\phi^2}
   \lambda ^2+4 c_{\phi^4}^{(12)} \lambda -4 c_{\phi^4}^{(4)} \lambda -2
   c_{\phi^4}^{(6)} \lambda \, \nonumber \\
&+\frac{3}{2}
   c_{W^2D^4}  g_1^2  g_2^2+\frac{5 c_{W^2D^4}
    g_2^4}{4}-\frac{5 c_{W^2\phi^2D^2}^{(11)}  g_2^2}{4}+\frac{5
   c_{W^2\phi^2D^2}^{(13)}  g_2^2}{4}+\frac{5 c_{W^2\phi^2D^2}^{(19)}
    g_2^2}{2}\, \nonumber \\
&+\frac{5 c_{W^3D^2}^{(1)}  g_2^3}{2}-\frac{5
   c_{W^3D^2}^{(2)}  g_2^3}{4}+\frac{7 c_{WB\phi^2D^2}^{(10)}
    g_1  g_2}{4}+\frac{3 c_{WB\phi^2D^2}^{(11)}  g_1
    g_2}{4}-\frac{3 c_{WB\phi^2D^2}^{(13)}  g_1
    g_2}{2}\, \nonumber \\
&+\frac{5 c_{WB\phi^2D^2}^{(8)}  g_1  g_2}{4}-5
   c_{W\phi^2D^4}^{(1)}  g_2 \lambda -\frac{3}{4} c_{W\phi^2D^4}^{(3)}
    g_1^2  g_2+5 c_{W\phi^2D^4}^{(3)}  g_2 \lambda
   -\frac{5 c_{W\phi^4D^2}^{(6)}  g_2}{2}\, \nonumber \\
&-\frac{3 c_{W\phi^4D^2}^{(7)}
    g_2}{2}\,,\\
 c_{\phi^6}^{(2)} &\rightarrow c_{\phi^6}^{(2)} -\frac{1}{4} c_{B^2D^4}  g_1^2
    g_2^2-\frac{c_{B^2\phi^2D^2}^{(8)}  g_1^2}{2}-2
   c_{B\phi^2D^4}^{(1)}  g_1 \lambda +\frac{1}{8} c_{B\phi^2D^4}^{(3)}
    g_1  g_2^2+2 c_{B\phi^2D^4}^{(3)}  g_1 \lambda\, \nonumber \\
&   -c_{B\phi^4D^2}^{(3)}  g_1+c_{\phi^2}  g_1^2 \lambda +2
   c_{\phi^4}^{(12)} \lambda -2 c_{\phi^4}^{(6)} \lambda
   +c_{W^2D^4}  g_1^2
    g_2^2-\frac{c_{WB\phi^2D^2}^{(10)}  g_1
    g_2}{2} \, \nonumber \\
&-\frac{3 c_{WB\phi^2D^2}^{(11)}  g_1
    g_2}{4}-c_{WB\phi^2D^2}^{(13)}  g_1  g_2-\frac{1}{2}
   c_{W\phi^2D^4}^{(3)}  g_1^2  g_2+\frac{c_{W\phi^4D^2}^{(7)}
    g_2}{4}\,,\\
 c_{\phi^4}^{(1)} &\rightarrow  c_{\phi^4}^{(1)} + c_{B^2D^4} g_1^2- c_{B\phi^2D^4}^{(3)} g_1 - c_{W^2D^4} g_2^2+ c_{W\phi^2D^4}^{(3)} g_2 \,,\\
 c_{\phi^4}^{(2)} &\rightarrow  c_{\phi^4}^{(2)} - c_{B^2D^4} g_1^2 + c_{B\phi^2D^4}^{(3)} g_1 - c_{W^2D^4} g_2^2+ c_{W\phi^2D^4}^{(3)} g_2\,,\\
 c_{\phi^4}^{(3)} &\rightarrow c_{\phi^4}^{(3)} +2 c_{W^2D^4} g_2^2 - 2 c_{W\phi^2D^4}^{(3)} g_2 \,,\\
 c_{G^3\phi^2}^{(1)} &\rightarrow  c_{G^3\phi^2}^{(1)} + g_3c_{G^2\phi^2D^2}^{(6)} \,,\\
 %
 %
 c_{W^3\phi^2}^{(1)} &\rightarrow c_{W^3\phi^2}^{(1)} - \frac{c_{W^3D^2}^{(1)} g_2^2}{2}\,,\\
 c_{W^3\phi^2}^{(2)} &\rightarrow c_{W^3\phi^2}^{(2)} - \frac{c_{W^3D^2}^{(3)} g_2^2}{2}\,,\\
 c_{W^2B\phi^2}^{(1)} &\rightarrow c_{W^2B\phi^2}^{(1)} -\frac{c_{W^3D^2}^{(1)} g_1
   g_2}{2}+\frac{c_{WB\phi^2D^2}^{(11)} g_2 }{2}+c_{WB\phi^2D^2}^{(13)} g_2 \,,\\
 c_{W^2B\phi^2}^{(2)} &\rightarrow c_{W^2B\phi^2}^{(2)} +\frac{c_{W^3D^2}^{(3)} g_1
   g_2}{4}+c_{WB\phi^2D^2}^{(19)} g_2+\frac{c_{WB\phi^2D^2}^{(17)} g_2}{2}\,,\\
 c_{G^2\phi^4}^{(1)} &\rightarrow c_{G^2\phi^4}^{(1)} + \lambda c_{G^2\phi^2D^2}^{(4)}\,,\\
 c_{G^2\phi^4}^{(2)} &\rightarrow  c_{G^2\phi^4}^{(2)} -4\lambda c_{G^2\phi^2D^2}^{(9)} \,,\\
 c_{W^2\phi^4}^{(1)} &\rightarrow c_{W^2\phi^4}^{(1)} -\frac{1}{8} c_{B^2D^4} g_1^2
   g_2^2+\frac{1}{16} c_{B\phi^2D^4}^{(3)} g_1
   g_2^2-\frac{c_{W^2D^4}
   g_2^4}{8}+\frac{c_{W^2\phi^2D^2}^{(11)} g_2^2}{4}\,\\
& +2
   c_{W^2\phi^2D^2}^{(14)} \lambda -\frac{c_{W^2\phi^2D^2}^{(19)}
   g_2^2}{2}-\frac{c_{W^3D^2}^{(1)}
   g_2^3}{2}+\frac{c_{W^3D^2}^{(2)}
   g_2^3}{4}-\frac{c_{WB\phi^2D^2}^{(11)} g_1
   g_2}{8} \, \\
& -\frac{c_{WB\phi^2D^2}^{(8)} g_1
   g_2}{4}+\frac{c_{W\phi^4D^2}^{(7)} g_2}{8} -\frac{1}{2} c_{\phi^2}g_2^2 \lambda\,, \\
 c_{W^2\phi^4}^{(2)} &\rightarrow  c_{W^2\phi^4}^{(2)} + \frac{c_{W^2\phi^2D^2}^{(10)} g_2^2}{4}+2
   c_{W^2\phi^2D^2}^{(16)} \lambda -\frac{c_{W^2\phi^2 D^2}^{(7)}
   g_2^2}{2}+\frac{c_{W^3D^2}^{(3)}
   g_2^3}{2}+\frac{c_{W^3D^2}^{(4)}
   g_2^3}{4}\,\\
&+\frac{c_{WB\phi^2D^2}^{(16)} g_1
   g_2}{4}+\frac{c_{WB\phi^2D^2}^{(19)} g_1
   g_2}{16} \,, \\
 c_{W^2\phi^4}^{(3)} &\rightarrow  c_{W^2\phi^4}^{(3)} + \frac{1}{16} c_{B\phi^2D^4}^{(3)} g_1
   g_2^2+\frac{c_{W\phi^4D^2}^{(7)}
   g_2}{8} \,,\\
 c_{W^2\phi^4}^{(4)} &\rightarrow  c_{W^2\phi^4}^{(4)}-\frac{c_{WB\phi^2D^2}^{(16)} g_1
   g_2}{4}-\frac{c_{WB\phi^2D^2}^{(19)} g_1
   g_2}{16}\,,\\
 c_{WB\phi^4}^{(1)} &\rightarrow c_{WB\phi^4}^{(1)} - \frac{c_{B^2\phi^2D^2}^{(6)} g_1
   g_2}{4}+\frac{c_{B^2\phi^2D^2}^{(8)} g_1
   g_2}{4}+\frac{1}{8} c_{B\phi^2D^4}^{(3)} g_1^2
   g_2-\frac{1}{2} c_{W^2D^4} g_1
   g_2^3 \, \nonumber \\
&+\frac{c_{W^2\phi^2D^2}^{(11)} g_1
   g_2}{4}-\frac{c_{W^2\phi^2D^2}^{(19)} g_1
   g_2}{2}-\frac{1}{2} c_{W^3D^2}^{(1)} g_1
   g_2^2+\frac{1}{4} c_{W^3D^2}^{(2)} g_1
   g_2^2-\frac{c_{WB\phi^2D^2}^{(11)}
   g_1^2}{8}\, \nonumber \\
&+\frac{c_{WB\phi^2D^2}^{(11)}
   g_2^2}{8}+c_{WB\phi^2D^2}^{(11)} \lambda
   +\frac{c_{WB\phi^2D^2}^{(13)}
   g_2^2}{4}-\frac{c_{WB\phi^2D^2}^{(8)}
   g_1^2}{4}+\frac{1}{4} c_{W\phi^2D^4}^{(3)}
   g_1 g_2^2\, \nonumber \\
&+\frac{c_{W\phi^4D^2}^{(7)} g_1}{4} -c_{\phi^2} g_1g_2 \lambda\,,\\
 c_{WB\phi^4}^{(2)} &\rightarrow c_{WB\phi^4}^{(2)}+ \frac{c_{B^2\phi^2 D^2}^{(11)} g_1
   g_2}{4}+\frac{c_{W^2\phi^2D^2}^{(10)} g_1
   g_2}{4}-\frac{c_{W^2\phi^2 D^2}^{(7)} g_1
  g_2}{2}+\frac{1}{2} c_{W^3D^2}^{(3)} g_1
   g_2^2\,\nonumber \\
&+\frac{1}{4} c_{W^3D^2}^{(4)} g_1
   g_2^2-\frac{c_{WB\phi^2D^2}^{(19)}
   g_2^2}{4}+c_{WB\phi^2D^2}^{(19)} \lambda
   -\frac{c_{WB\phi^2D^2}^{(17)} g_2^2}{4}\,,\\
 c_{B^2\phi^4}^{(1)} &\rightarrow c_{B^2\phi^4}^{(1)} + \frac{c_{B^2D^4} g_1^4}{8}+c_{B^2\phi^2D^2}^{(4)}
   \lambda -\frac{c_{B^2\phi^2D^2}^{(6)}
   g_1^2}{4}+\frac{c_{B^2\phi^2D^2}^{(8)}
   g_1^2}{4}-\frac{3}{8}
   c_{W^2D^4} g_1^2
   g_2^2\, \nonumber \\
&+\frac{c_{WB\phi^2D^2}^{(11)} g_1
   g_2}{8}+\frac{c_{WB\phi^2D^2}^{(13)} g_1
   g_2}{4}+\frac{1}{4} c_{W\phi^2D^4}^{(3)} g_1^2
   g_2 -\frac{1}{2}c_{\phi^2} g_1^2 \lambda\, , \\
 c_{B^2\phi^4}^{(2)} &\rightarrow c_{B^2\phi^4}^{(2)}+ \frac{c_{B^2\phi^2 D^2}^{(11)} g_1^2}{4}-4 c_{B^2\phi^2D^2}^{(9)}
   \lambda -\frac{c_{WB\phi^2D^2}^{(19)}
   g_1 g_2}{4}-\frac{c_{WB\phi^ 2D^2}^{(23)} g_1
   g_2}{4}\,,\\
 %
 %
 c_{G^2\phi^2 D^2}^{(2)} &\rightarrow c_{G^2\phi^2 D^2}^{(2)} - \frac{1}{2} c_{G^2\phi^2 D^2}^{(4)}\,,\\
 %
 %
 %
 c_{W^2\phi^2 D^2}^{(2)} &\rightarrow c_{W^2\phi^2 D^2}^{(2)} - c_{W^2\phi^2 D^2}^{(14)} \,,\\
 c_{W^2\phi^2 D^2}^{(3)} &\rightarrow  c_{W^2\phi^2 D^2}^{(3)} -  c_{W^2\phi^2 D^2}^{(16)}\,,\\
 c_{W^2\phi^2 D^2}^{(4)} &\rightarrow c_{W^2\phi^2 D^2}^{(4)} -2 c_{W^3D^2}^{(1)} g_2\,,\\
 %
 %
 c_{W^2\phi^2 D^2}^{(6)} &\rightarrow  c_{W^2\phi^2 D^2}^{(6)}-c_{W^3D^2}^{(3)} g_2\,,\\
 c_{WB\phi^2 D^2}^{(1)} &\rightarrow c_{WB\phi^2 D^2}^{(1)} - \frac{c_{WB\phi^2D^2}^{(11)}}{2}\,,\\
 c_{WB\phi^2 D^2}^{(2)} &\rightarrow c_{WB\phi^2 D^2}^{(2)}-\frac{c_{WB\phi^2D^2}^{(19)}}{2}\,,\\
 c_{WB\phi^2 D^2}^{(3)} &\rightarrow c_{WB\phi^2 D^2}^{(3)} - c_{W^2BD^2}^{(1)} g_2+c_{W^2BD^2}^{(2)}
   g_2+c_{WB\phi^2D^2}^{(12)}+2
   c_{WB\phi^2D^2}^{(7)}\,,\\
 %
 %
 c_{WB\phi^2 D^2}^{(5)} &\rightarrow c_{WB\phi^2 D^2}^{(5)} -c_{W^2BD^2}^{(3)} g_2+\frac{3 c_{W^2BD^2}^{(4)}
   g_2}{2}+2 c_{WB\phi^2D^2}^{(15)}+c_{WB\phi^2D^2}^{(18)}\,,\\
 %
 %
 %
 c_{B^2\phi^2 D^2}^{(2)} &\rightarrow c_{B^2\phi^2 D^2}^{(2)}-\frac{c_{B^2\phi^2D^2}^{(4)}}{2}\,,\\
 %
 %
 c_{W\phi^4 D^2}^{(1)} &\rightarrow c_{W\phi^4 D^2}^{(1)} - c_{B^2D^4} g_1^2 g_2+\frac{c_{B\phi^2D^4}^{(3)}
   g_1 g_2}{2}-c_{W^2D^4}
   g_2^3+c_{W^2\phi^2D^2}^{(11)} g_2-4
   c_{W^2\phi^2D^2}^{(19)} g_2 \, \nonumber \\
& -4 c_{W^3D^2}^{(1)}
   g_2^2+2 c_{W^3D^2}^{(2)} g_2^2-\frac{3
   c_{WB\phi^2D^2}^{(11)} g_1}{2}-3 c_{WB\phi^2D^2}^{(8)}
   g_1-\frac{c_{W\phi^2D^4}^{(3)}
   g_2^2}{2} \, \nonumber \\
&+2 c_{W\phi^4D^2}^{(7)}\,,\\
 c_{W\phi^4 D^2}^{(2)} &\rightarrow c_{W\phi^4 D^2}^{(2)}+ c_{W^2\phi^2D^2}^{(10)} g_2-4 c_{W^2\phi^2 D^2}^{(7)}
   g_2+4 c_{W^3D^2}^{(3)} g_2^2+2 c_{W^3D^2}^{(4)}
   g_2^2+2 g_1 c_{WB\phi^2D^2}^{(16)}\, \nonumber \\
& +\frac{c_{WB\phi^2D^2}^{(19)}
   g_1}{2}\,,\\
 c_{W\phi^4 D^2}^{(3)} &\rightarrow c_{W\phi^4 D^2}^{(3)}+\frac{c_{W^2BD^2}^{(1)} g_1
   g_2}{2}+c_{W^2\phi^2D^2}^{(12)}
   g_2+c_{W^2\phi^2D^2}^{(18)}
   g_2-\frac{c_{WB\phi^2D^2}^{(12)}
   g_1}{2}-c_{WB\phi^2D^2}^{(7)}
   g_1 \, \nonumber \\
&-c_{WB\phi^2D^2}^{(9)} g_1\,,\\
 c_{W\phi^4 D^2}^{(4)} &\rightarrow c_{W\phi^4 D^2}^{(4)} + \frac{c_{W^2BD^2}^{(3)} g_1
   g_2}{2}+\frac{c_{W^2BD^2}^{(4)} g_1
   g_2}{4}-c_{W^2\phi^2 D^2}^{(8)}
   g_2+c_{W^2\phi^2D^2}^{(9)} g_2+c_{WB\phi^2D^2}^{(14)}
   g_1\, \nonumber \\
&-c_{WB\phi^2D^2}^{(15)}
   g_1-\frac{c_{WB\phi^2D^2}^{(18)}
   g_1}{2}\,,\\
 c_{B\phi^4 D^2}^{(1)} &\rightarrow c_{B\phi^4 D^2}^{(1)} +  c_{B^2D^4} g_1^3-c_{B^2\phi^2D^2}^{(6)}
   g_1+c_{B^2\phi^2D^2}^{(8)}
   g_1-\frac{c_{B\phi^2D^4}^{(3)}
   g_1^2}{2}-3 c_{W^2D^4}
   g_1 g_2^2 \, \nonumber \\
&+\frac{3 c_{WB\phi^2D^2}^{(11)}
   g_2}{2}+3 c_{WB\phi^2D^2}^{(13)} g_2+\frac{3
   c_{W\phi^2D^4}^{(3)} g_1 g_2}{2}\,,\\\label{eq:reductionF}
 c_{B\phi^4 D^2}^{(2)} &\rightarrow c_{B\phi^4 D^2}^{(2)} + c_{B^2\phi^2 D^2}^{(11)} g_1-3
   c_{WB\phi^2D^2}^{(19)} g_2-3 c_{WB\phi^2D^2}^{(17)}
   g_2\,.
\end{align}
Wilson coefficients absent in the equations above are not modified by redundant interactions.

This concludes the reduction of the redundant operators to the physical basis. One more important remark is in order, though. The physical operators get also corrections from pairs of redundant dimension-six interactions that, as we focus only on the dimension-eight sector, we do not specify. In general, such corrections can not be derived by applying the dimension-six SMEFT equations of motion on the redundant dimension-six terms; the latter must instead be removed by proper field redefinitions~\cite{Criado:2018sdb}. This exercise will be addressed elsewhere. 

\section{Some applications}
\label{sec:applications}

\subsection{Integrating out a scalar singlet to one loop}
Let us extend the SM with a heavy singlet scalar $\mathcal{S}\sim (1,1)_0$. The numbers within parentheses and the sub-index indicate the $SU(3)_c$ and $SU(2)_L$ quantum numbers and the hypercharge, respectively. We assume a $\mathbb{Z}_2$ symmetry $\mathcal{S}\to -\mathcal{S}$, so that the new physics Lagrangian reads:
\begin{align}
 \mathcal{L}_\text{NP} = \frac{1}{2} (D_\mu\mathcal{S})( D^\mu\mathcal{S})-\frac{1}{2}m_\mathcal{S}^2\mathcal{S}^2 
 -\lambda_{\mathcal{S}\phi}\mathcal{S}^2\phi^\dagger\phi
 -\lambda_\mathcal{S}\mathcal{S}^4\,.
\end{align}
Because of the $\mathbb{Z}_2$ symmetry, all effective operators arise first at one loop. Hence, the contribution from redundant dimension-six interactions to the dimension-eight terms upon field redefinitions are formally two-loop corrections and therefore negligible within our order of calculation. Consequently, Eqs.~\eqref{eq:reductionI}-\eqref{eq:reductionF} are valid without any further corrections.

Thus, we implement this model, together with our Green's basis of operators (including the on-shell relations in Eqs.~\eqref{eq:reductionI}-\eqref{eq:reductionF}) in \texttt{matchmakereft}~\cite{Carmona:2021xtq}. Automatically, we get the following dimension-eight Wilson coefficients:
\begin{align}
 \frac{c_{\phi^6}^{(1)}}{\Lambda^4} &= \frac{1}{1920 \,m_\mathcal{S}^4\,\pi^2} \lambda_{\mathcal{S}\phi}^2 \left(5\lambda_{\mathcal{S}\phi} -8\lambda\right) \,, \\
\frac{c_{\phi^4}^{(3)}}{\Lambda^4} &= \frac{1}{960\, m_\mathcal{S}^4\,\pi^2} \lambda_{\mathcal{S}\phi}^2\,. 
\end{align}
For simplicity, we have taken the limit $g_2\rightarrow 0$. (Also, we have not computed the Wilson coefficient of $\mathcal{O}_{\phi^8}$.)
To the best of our knowledge, this is the first one-loop computation of the matching of a scalar singlet onto the bosonic SMEFT to dimension-eight.

\subsection{Integrating out a scalar quadruplet to one loop}
\label{sec:quadruplet}

In this case, we consider the SM extended with a scalar $SU(2)_L$ quadruplet with $Y=1/2$ and mass $m_\Theta$. We name it as $\Theta$. The relevant new physics Lagrangian is:
\begin{align}
 \mathcal{L}_{\text{NP}} &= D_\mu\Theta^\dagger D^\mu\Theta-m_\Theta^2\Theta^\dagger\Theta-\lambda_{\Theta} (\phi^\dagger\sigma^I\phi) C^\alpha_{I\beta}\tilde{\phi}^\beta\epsilon_{\alpha \gamma} \Theta^\gamma +\text{h.c.}\,.
\end{align}
(For simplicity, we are ignoring other quartic terms.) The $C^\alpha_{I\beta}$ symbol represents the Clebsh-Gordan needed to single out the $SU(2)_L$ singlet from the contraction of a quadruplet, a doublet and a triplet.
At tree level and dimension six, only the operator $(\phi^\dagger\phi)^3$ is generated. (Therefore, once more, indirect contributions to the dimension-eight Wilson coefficients can be obtained simply from Eqs.~\eqref{eq:reductionI}-\eqref{eq:reductionF}.) At dimension eight, and again in the limit $g_2\to 0$, we obtain:
\begin{align}
 \frac{c_{B^4}^{(1)}}{\Lambda^4} &=\frac{7g_1^4}{92160\, m_\Theta^4\,\pi^2}\,, \\
\frac{c_{B^4}^{(2)}}{\Lambda^4} &= \frac{g_1^4}{92160\, m_\Theta^4\,\pi^2} \,, \\
\frac{c_{\phi^6}^{(1)}}{\Lambda^4} &= \frac{|\lambda_\Theta|^2}{3 \,m_\Theta^2}+\frac{ - 6440 \,g_1^2\, |\lambda_\Theta|^2+103040\, |\lambda_\Theta|^2 
   \lambda}{80640 \, m_\Theta^4\, \pi ^2} \,, \\
\frac{c_{\phi^6}^{(2)}}{\Lambda^4} &= -\frac{|\lambda_\Theta|^2}{2 \,m_\Theta^2}+\frac{+3640\, g_1^2 |\lambda_\Theta|^2 \,- 655200\, |\lambda_\Theta|^2\,\lambda  }{483840 \, m_\Theta^4 \,\pi^2} \,, \\
\frac{c_{\phi^4}^{(1)}}{\Lambda^4} &= \frac{4480 \,|\lambda_\Theta|^2- 3 g_1^4}{40320
   \, m_\Theta^4 \,\pi^2} \,, \\
\frac{c_{\phi^4}^{(2)}}{\Lambda^4} &=\frac{3 g_1^4+1120 \,|\lambda_\Theta|^2}{40320
   \, m_\Theta^4 \,\pi^2} \,, \\
\frac{c_{\phi^4}^{(3)}}{\Lambda^4} &= -\frac{|\lambda_\Theta|^2}{18 \, m_\Theta^4 \,\pi^2} \,, \\
\frac{c_{B^2\phi^4}^{(1)}}{\Lambda^4} &= \frac{1960\, g_1^2 |\lambda_\Theta|^2-3
   g_1^6}{322560 \, m_\Theta^4 \,\pi^2} \,, \\
\frac{c_{B\phi^4D^2}^{(1)}}{\Lambda^4} &= -\frac{g_1^5}{13440 \, m_\Theta^4 \,\pi^2}\,.
\end{align}
The tree-level contribution to $c_{\phi^6}^{(1)}$ and $c_{\phi^6}^{(2)}$ had been previously computed in Ref.~\cite{Murphy:2020rsh}, and we agree with the result therein. We are also in agreement with the (loop produced) $c_{B^2}^{(1)}$ and $c_{B^2}^{(2)}$ previously reported in Refs.~\cite{Quevillon:2018mfl,Remmen:2019cyz}. 

\subsection{Reduction of Lagrangian to a physical basis}
Functional methods comprise a very powerful tool to match UV models onto EFTs, by literally \textit{integrating} over the heavy dynamical fields in the path integral~\cite{Gaillard:1985uh,Cheyette:1987qz,Henning:2014wua}. The advantage of this approach with respect to matching 1PI amplitudes in Feynman diagrams is that no basis of EFT operators (neither Green's nor physical) must be known in advance to complete the calculation. 

The drawback, though, is that the resulting EFT Lagrangian is highly redundant, involving operators related by equations of motion, integration by parts and algebraic identities. Operators can be even connected by a re-labeling of dummy indices. (However simple this might look, they can be hard to differentiate in a more or less automatic fashion.) 

With the help of \texttt{SuperTracer}~\cite{Fuentes-Martin:2020udw}, we have checked the \textit{monstruosity} of the EFT Lagrangian resulting from integrating out heavy fields to one loop in elaborated models. However, the cumbersone mixture of EFT interactions can be already appreciated in simple models and even at tree level.

Let us consider, for example, an extension of the SM with a heavy real vector triplet $\mathcal{W}\sim (1,3)_0$, with Lagrangian:
\begin{align}
 \mathcal{L}_\text{NP} = \frac{1}{2} \bigg[D_{\mu} \mathcal{W}_{\nu}^\dagger D^\nu\mathcal{W}^\mu-D_\mu\mathcal{W}^\dagger_\nu D^\mu \mathcal{W}^\nu+ m_\mathcal{W}^2 \mathcal{W}_\mu^\dagger\mathcal{W}^\mu + (g_\mathcal{W}^\phi \mathcal{W}^\mu\phi^{I\dagger}\sigma^I i D_\mu\phi +\text{h.c.})\bigg]\,.
\end{align}
At tree level, the effective action is given by the UV action evaluated on $\mathcal{W} = \mathcal{W}_c$, namely the classical configuration that solves the $\mathcal{W}$ equations of motion. We compute this to order $1/m_{\mathcal{W}}^4$ using \texttt{MatchingTools}~\cite{Criado:2017khh}, obtaining:
\begin{align}\nonumber
 \mathcal{L}^{(8)}_\text{EFT} &= \frac{(g_\mathcal{W}^\phi)^2}{m_\mathcal{W}^4}\bigg[2 (D_\mu\phi^\dagger D_\nu\phi) (D^\mu\phi^\dagger D^\nu\phi) + 4 (D_\nu\phi^\dagger D^\nu D^\mu\phi) (D_\mu\phi^\dagger\phi) - 2 (D_\mu\phi^\dagger D_\nu\phi) (\phi^\dagger D^\mu D^\nu\phi)\\\nonumber
 &-4 (D_\mu\phi^\dagger\phi) (D^\mu D_\nu\phi^\dagger D^\nu\phi) +2 (D_\mu\phi^\dagger D_\nu\phi) (D^\mu D^\nu \phi^\dagger \phi)-4 (D_\mu\phi^\dagger D^\mu\phi) (D_\nu\phi^\dagger D^\nu\phi) \\\nonumber
 &+ 2 (D_\mu\phi^\dagger D_\nu\phi)(D^\nu \phi^\dagger D^\mu\phi) + \frac{1}{2}(\phi^\dagger D_\mu D_\nu\phi) (\phi^\dagger D^\mu D^\nu\phi) - 2 (D_\nu D_\rho \phi^\dagger D^\nu D^\rho\phi) (\phi^\dagger\phi) \\\nonumber
 & + (D_\mu D_\nu \phi^\dagger \phi)( \phi^\dagger D^\mu D^\nu\phi) - 4(\phi^\dagger D_\rho\phi) (D_\nu\phi^\dagger D^\rho D^\nu\phi) + 2(\phi^\dagger D_\nu D_\mu\phi) (D^\mu\phi^\dagger D^\nu\phi)\\\nonumber
 &+\frac{1}{2} (D_\mu D_\nu\phi^\dagger\phi) (D^\mu D^\nu\phi^\dagger\phi)+4 (D_\rho D_\nu\phi^\dagger D^\rho\phi)(D^\nu\phi^\dagger\phi) - 2(D_\nu D_\mu\phi^\dagger\phi)(D^\mu\phi^\dagger D^\nu\phi)\\\nonumber
 &-\frac{1}{2}(\phi^\dagger D_\nu D_\mu\phi)(\phi^\dagger D^\mu D^\nu\phi) + 2(D_\rho D_\nu\phi^\dagger D^\nu D^\rho\phi)(\phi^\dagger\phi)-(D^\nu D^\mu\phi^\dagger\phi)(\phi^\dagger D_\mu D_\nu\phi)\\\nonumber
 &-\frac{1}{2}(D_\nu D_\mu\phi^\dagger\phi)(D^\mu D^\nu\phi^\dagger\phi)\bigg]\,.
\end{align}
(The dimension-six piece can be checked in Ref.~\cite{deBlas:2017xtg}, from where we borrow notation.) We reproduce precisely the different dummy indices resulting from the automatic calculation. It is apparent that this Lagrangian can not easily (at least not immediately) be reduced to a physical basis by hand.

However, one could simply (and automatically) export $\mathcal{L}_\text{EFT}^{(8)}$ to \texttt{FeynArts} with the help of \texttt{FeynRules}, compute the relevant 1PI tree-level off-shell amplitudes and project the results onto our basis.~\footnote{\texttt{Matchmakereft} includes also a single instruction to perform this action automatically.} Proceeding this way, we obtain:
\begin{align}\nonumber
 \mathcal{L}_\text{EFT}^{(8)} = \frac{(g_\mathcal{W}^\phi)^2}{m_\mathcal{W}^4} \bigg[2\mathcal{O}_{\phi^{4}}^{(1)} &+ 2\mathcal{O}_{\phi^{4}}^{(2)}-4\mathcal{O}_{\phi^{4}}^{(3)} -\frac{1}{4} g_2^2\mathcal{O}_{W^2\phi^4}^{(1)}+\frac{1}{2}g_1 g_2\mathcal{O}_{WB\phi^4}^{(1)}\\
 &+\frac{3}{4}g_1^2\mathcal{O}_{B^2\phi^4}^{(1)}-2g_2\mathcal{O}_{W\phi^4 D^2}^{(1)}+6g_1\mathcal{O}_{B\phi^4 D^2}^{(1)}+2g_1\mathcal{O}_{B\phi^4 D^2}^{(3)}\bigg]\,.
\end{align}

Thus, even matching computations performed using functional methods can benefit from knowing a basis of independent Green's functions.

\section{Conclusions and future directions}
\label{sec:conclusions}
Off-shell calculations are common practice within effective quantum field theories. They have the advantage that they involve a substantially smaller amount of Feynman diagrams than calculations of physical (on-shell) S-matrix elements. In contrast, they require introducing redundant interactions in the Lagrangian, including nonphysical terms that vanish under field redefinitions.

Concentrating on the SMEFT, it would then be desirable to have a complete set of operators independent off shell, so neither related by algebraic identities nor by integration by parts. Any such set of interactions is called a \textit{Green's basis}~\cite{Jiang:2018pbd}; a particular realisation to dimension six was built in Ref.~\cite{Gherardi:2020det}. In this paper, we have constructed the bosonic dimension-eight counterpart, which consists of $86$ new interactions.

One important aspect of our approach has been working in momentum space to establish the off-shell independence of operators, thus avoiding the otherwise cumbersome operations needed at the level of the Lagrangian when the interactions involve many fields and derivatives. In particular, integration by parts amounts simply to removing one momentum in 1PI amplitudes. Other relations, such as four-dimensional constraints resulting from contractions of the Levi-Civita symbol, are harder to enforce systematically at the level of amplitudes, but we have shown that they can be accounted for by requiring that at most four Lorentz vectors (momenta or polarisations) are linearly independent in four dimensions.

Our Green's basis is obviously not unique. Infinitely many other combinations of operators could be considered, in particular bases in which the redundant interactions are related to physical ones by equations of motion through simpler relations than those in Eqs.~\eqref{eq:reductionI}--\eqref{eq:reductionF}. One advantage of the one presented here, though, is that the renormalisation of the $X^3\phi^2$ operators (which will be presented elsewhere~\cite{Bakshi:2021abc}) can be carried out without necessarily projecting the contractions of the Levi-Civita symbol onto four dimensions. This simplifies notably the calculation, because these operators generate amplitudes with three polarization vectors and five different momenta, from which three vectors must be therefore projected onto the remaining four independent ones.
(A more mundane albeit technically important property of our basis is that all the effective operators in there can be exported to \texttt{FeynArts} using \texttt{FeynRules}, which is not always the case.)

Renormalising the bosonic sector of the SMEFT to dimension eight (thus concluding the effort initiated in Ref.~\cite{Chala:2021pll}) is in fact an avenue we have already started to explore on the basis of this work. Another future direction of our current work includes classifying all independent bosonic evanescent operators (and projecting them onto the physical basis in four dimensions). This will simplify the matching of UV models onto the SMEFT using tools based on diagrammatic calculations, since the Levi-Civita symbol can be assumed to be simply a totally anti-symmetric tensor without further structure inherited from the four-dimensional space-time. (In fact, for performing the matching in Section~\ref{sec:quadruplet}, we augmented the basis with the operator $\mathcal{O}=B^{\mu\nu} B_{\nu\rho} B^{\rho\sigma} B_{\sigma\mu}$, which in $D=4$ fullfills $\mathcal{O} = \frac{1}{2}\mathcal{O}_{B^4}^{(1)}+ \frac{1}{4}\mathcal{O}_{B^4}^{(2)}$.)

Connected to this, one more avenue that we aim to pursue in the near future is using \texttt{matchmakereft} to analyse positivity bounds on SMEFT $X^2\phi^2 D^2$ operators (first presented in Ref.~\cite{Bi:2019phv}) in models in which low-momentum $2\to 2$ amplitudes are not necessarily well approximated by the EFT at tree level. Such models involve in general heavy fields with linear interactions, which can be integrated out only if loops involving heavy-light particles are calculated. To the best of our knowledge, other tools for matching, such as \texttt{Codex}~\cite{DasBakshi:2018vni}, do not include yet these loops, particularly for dimension-eight computations.

\section*{Acknowledgments}
We are grateful to Jose Santiago for sharing (and for instructing us on) \texttt{matchmakereft}, as well as for discussions and for comments on the manuscript. We are also enormously thankful to Renato Fonseca for the many enlightening discussions and for help with \texttt{Sym2Int}. MC would also like to thank Christopher Murphy for clarifying some calculations in Ref.~\cite{Murphy:2020rsh}.
MC and ADC are supported by SRA under grant 
PID2019-106087GB-C21 (10.13039/501100011033). MC is also supported by the Junta de Andaluc\'ia grants FQM 101, A-FQM-211-UGR18 and P18-FR-4314 (FEDER), as well as by the Spanish MINECO under the Ram\'on y Cajal programme. ADC is also supported by the Spanish MINECO under the FPI programme.
GG acknowledges support by LIP 
(FCT, COMPETE2020-Portugal2020, FEDER, POCI-01-0145-FEDER-007334) as well as by FCT under project CERN/FIS-PAR/0024/2019 and under the grant SFRH/BD/144244/2019.
\newpage
\appendix
\section{Tables of operators}
\label{app:tables}

\begin{table}[h!]
 \begin{center}
  \resizebox{0.97\textwidth}{!}{
\begin{tabular}{cclcl}
   \toprule\\[-0.3cm]
   & \textbf{Operator} & \textbf{Notation} & \textbf{Operator} & \textbf{Notation}\\[0.5cm]
   \rotatebox[origin=c]{90}{\boldmath{$\phi^8$}} & $( \phi^\dagger\phi)^4$ & $\mathcal{O}_{\phi^8}$ & &\\[0.1cm]
   \hline\\[-0.3cm]
    \multirow{2}{*}{\rotatebox[origin=c]{90}{\boldmath{$\phi^6 D^2$}}} &  $(\phi^{\dag} \phi)^2 (D_{\mu} \phi^{\dag} D^{\mu} \phi)$ & $\mathcal{O}_{\phi^6}^{(1)}$  &  $(\phi^{\dag} \phi) (\phi^{\dag} \sigma^I \phi) (D_{\mu} \phi^{\dag} \sigma^I D^{\mu} \phi)$ &  
$\mathcal{O}_{\phi^6}^{(2)}$ \\[0.2cm]
& \textcolor{gray}{$(\phi^\dagger\phi)^2 (\phi^\dagger D^2\phi + \text{h.c.})$} &\textcolor{gray}{ $\mathcal{O}_{\phi^6}^{(3)}$} &\textcolor{gray}{$(\phi^\dagger\phi)^2 D_\mu(\phi^\dagger\ii \overleftrightarrow{D}^\mu\phi)$} &\textcolor{gray}{ $\mathcal{O}_{\phi^6}^{(4)}$} \\[0.2cm]
   \hline\\[-0.3cm]
   \multirow{10}{*}{\rotatebox[origin=c]{90}{\boldmath{$\phi^4 D^4$}}}  &  $(D_{\mu} \phi^{\dag} D_{\nu} \phi) (D^{\nu} \phi^{\dag} D^{\mu} \phi)$ &  $\mathcal{O}_{\phi^4}^{(1)}$ &
 $(D_{\mu} \phi^{\dag} D_{\nu} \phi) (D^{\mu} \phi^{\dag} D^{\nu} \phi)$ & $\mathcal{O}_{\phi^4}^{(2)}$\\[0.2cm]
& $(D^{\mu} \phi^{\dag} D_{\mu} \phi) (D^{\nu} \phi^{\dag} D_{\nu} \phi)$ & $\mathcal{O}_{\phi^4}^{(3)}$ & \textcolor{gray}  {$D_\mu\phi^\dagger D^\mu\phi(\phi^\dagger D^2\phi + \text{h.c.})$} & \textcolor{gray}  {$\mathcal{O}_{\phi^4}^{(4)}$}\\[0.2cm]

& \textcolor{gray}  {$D_\mu\phi^\dagger D^\mu\phi(\phi^\dagger \ii D^2\phi + \text{h.c.})$} & \textcolor{gray}  {$\mathcal{O}_{\phi^4}^{(5)} $} & \textcolor{gray}  {$(D_\mu\phi^\dagger \phi) (D^2\phi^\dagger D_\mu\phi) + \text{h.c.}$} & \textcolor{gray}  {$\mathcal{O}_{\phi^4}^{(6)}$}  \\[0.2cm]
& \textcolor{gray}  {$(D_\mu\phi^\dagger \phi) (D^2\phi^\dagger \ii D_\mu\phi) + \text{h.c.}$} & \textcolor{gray}  {$\mathcal{O}_{\phi^4}^{(7)}$}  & \textcolor{gray}  { $(D^2\phi^\dagger\phi) (D^2\phi^\dagger\phi)+\text{h.c.}$} & \textcolor{gray}  {$\mathcal{O}_{\phi^4}^{(8)}$}\\[0.2cm]
& \textcolor{gray}  {$(D^2\phi^\dagger\phi) (\ii D^2\phi^\dagger\phi)+\text{h.c.}$} & \textcolor{gray}  {$\mathcal{O}_{\phi^4}^{(9)}$}& \textcolor{gray}  {$(D^2\phi^\dagger D^2\phi) (\phi^\dagger\phi)$} & \textcolor{gray}  {$\mathcal{O}_{\phi^4}^{(10)}$}\\[0.2cm]
& \textcolor{gray}  {$(\phi^\dagger D^2\phi) (D^2\phi^\dagger\phi)$} & \textcolor{gray}  {$\mathcal{O}_{\phi^4}^{(11)}$} &\textcolor{gray}  {$(D_\mu\phi^\dagger \phi)(D^\mu\phi^\dagger D^2\phi) + \text{h.c.}$} & \textcolor{gray}  {$\mathcal{O}_{\phi^4}^{(12)}$} \\[0.2cm]
& \textcolor{gray}  {$(D_\mu\phi^\dagger \phi)(D^\mu\phi^\dagger \ii D^2\phi) + \text{h.c.}$} & \textcolor{gray}  {$\mathcal{O}_{\phi^4}^{(13)}$} 
\\[0.3cm]
   \hline\\[-0.3cm]
   \multirow{4}{*}{\rotatebox[origin=c]{90}{\boldmath{ {$X^3 \phi^2$}}}} &$f^{ABC} (\phi^\dag\phi)G_\mu^{A,\nu}G_\nu^{B,\rho}G_\rho^{C,\mu}$ & $\mathcal{O}_{G^3\phi^2}^{(1)}$ & $f^{ABC} (\phi^\dag\phi)G_\mu^{A,\nu}G_\nu^{B,\rho}\tilde{G}_\rho^{C,\mu}$ & $\mathcal{O}_{G^3\phi^2}^{(1)}$  \\[0.2cm] 
 & {$\epsilon^{IJK} (\phi^\dag \phi) W_{\mu}^{I\nu} W_{\nu}^{J\rho} W_{\rho}^{K\mu}$} &  {$\mathcal{O}_{W^3\phi^2}^{(1)}$} &  {$\epsilon^{IJK} (\phi^\dag \phi) W_{\mu}^{I\nu} W_{\nu}^{J\rho} \widetilde{W}_{\rho}^{K\mu}$} &  
 {$\mathcal{O}_{W^3\phi^2}^{(2)}$} \\[0.2cm]
&  {$\epsilon^{IJK} (\phi^\dag \sigma^I \phi) B_{\mu}^{\,\nu} W_{\nu}^{J\rho} W_{\rho}^{K\mu}$} &  {$\mathcal{O}_{W^2B\phi^2}^{(1)}$} &  {$\epsilon^{IJK} (\phi^\dag \sigma^I \phi) (\widetilde{B}^{\mu\nu} W_{\nu\rho}^J W_{\mu}^{K\rho} + B^{\mu\nu} W_{\nu\rho}^J \widetilde{W}_{\mu}^{K\rho})$} &  {$\mathcal{O}_{W^2B\phi^2}^{(2)}$}\\[0.2cm]
   \hline\\[-0.3cm]
   \multirow{6}{*}{\rotatebox[origin=c]{90}{\boldmath{$X^2 \phi^4$}}} & $(\phi^\dag \phi)^2 G_{\mu\nu}^A G^{A\mu\nu}$ & $O_{G^2\phi^4}^{(1)}$ & $(\phi^\dag \phi)^2 \widetilde{G}_{\mu\nu}^A G^{A\mu\nu}$ & $O_{G^2\phi^4}^{(2)}$\\[0.2cm]
   &$(\phi^\dag \phi)^2 W^I_{\mu\nu} W^{I\mu\nu}$ & $\mathcal{O}_{W^2\phi^4}^{(1)}$ &   $(\phi^\dag \phi)^2 \widetilde W^I_{\mu\nu} W^{I\mu\nu}$ & 
$\mathcal{O}_{W^2\phi^4}^{(2)}$\\[0.2cm]
& $(\phi^\dag \sigma^I \phi) (\phi^\dag \sigma^J \phi) W^I_{\mu\nu} W^{J\mu\nu}$ & $\mathcal{O}_{W^2\phi^4}^{(3)}$  &
$(\phi^\dag \sigma^I \phi) (\phi^\dag \sigma^J \phi) \widetilde W^I_{\mu\nu} W^{J\mu\nu}$ & $\mathcal{O}_{W^2\phi^4}^{(4)}$  \\[0.2cm]
& $ (\phi^\dag \phi) (\phi^\dag \sigma^I \phi) W^I_{\mu\nu} B^{\mu\nu}$ & $\mathcal{O}_{WB\phi^4}^{(1)}$ & $(\phi^\dag \phi) (\phi^\dag \sigma^I \phi) \widetilde W^I_{\mu\nu} B^{\mu\nu}$ & $\mathcal{O}_{WB\phi^4}^{(2)}$ \\[0.2cm]
& $ (\phi^\dag \phi)^2 B_{\mu\nu} B^{\mu\nu}$ & $\mathcal{O}_{B^2\phi^4}^{(1)}$ & $(\phi^\dag \phi)^2 \widetilde B_{\mu\nu} B^{\mu\nu}$ & $\mathcal{O}_{B^2\phi^4}^{(2)}$\\[0.4cm]
   \hline\\[-0.3cm]
   \multirow{6}{*}{\rotatebox[origin=c]{90}{ {\boldmath{$X \phi^2 D^4$}}}} &    \textcolor{gray}  {$i (D_\nu\phi^\dagger \sigma^I D^2\phi-D^2\phi^\dagger\sigma^I D_\nu\phi) D_\mu W^{I\mu\nu}$} &  \textcolor{gray}  {$ \mathcal{O}_{W\phi^2 D^4}^{(1)} $}  &
\textcolor{gray}   {$ (D_\nu\phi^\dagger \sigma^I D^2\phi+D^2\phi^\dagger\sigma^I D_\nu\phi) D_\mu W^{I\mu\nu}$} & \textcolor{gray}   {$\mathcal{O}_{W\phi^2 D^4}^{(2)} $} \\[0.2cm]
&  \textcolor{gray}   {$ i (D_\rho D_\nu\phi^\dagger\sigma^I D^\rho\phi-D^\rho\phi^\dagger \sigma^I D_\rho D_\nu\phi) D_\mu W^{I\mu\nu}$} &  \textcolor{gray}  {$\mathcal{O}_{W\phi^2 D^4}^{(3)}$} \\[0.2cm]
&   \textcolor{gray}  {$ i (D_\nu\phi^\dagger D^2\phi-D^2\phi^\dagger D_\nu\phi) D_\mu B^{\mu\nu}$} &  \textcolor{gray}{$\mathcal{O}_{B\phi^2 D^4}^{(1)}$} &   \textcolor{gray}{$(D_\nu\phi^\dagger D^2\phi+D^2\phi^\dagger D_\nu\phi) D_\mu B^{\mu\nu}$} &  \textcolor{gray}{$\mathcal{O}_{B\phi^2 D^4}^{(2)} $} \\[0.2cm]
&   \textcolor{gray}{$i (D_\rho D_\nu\phi^\dagger D^\rho\phi-D^\rho\phi^\dagger D_\rho D_\nu\phi) D_\mu B^{\mu\nu}$} &  \textcolor{gray}{$\mathcal{O}_{B\phi^2 D^4}^{(3)} $} \\[0.2cm]
   \hline\\[-0.3cm]
   \multirow{6}{*}{\rotatebox[origin=c]{90}{\boldmath{$X \phi^4 D^2$}}} & $\text{i}(\phi^{\dag} \phi) (D^{\mu} \phi^{\dag} \sigma^I D^{\nu} \phi) W_{\mu\nu}^I$ & $\mathcal{O}_{W\phi^4D^2}^{(1)}$  & $\text{i}(\phi^{\dag} \phi) (D^{\mu} \phi^{\dag} \sigma^I D^{\nu} \phi) \widetilde{W}_{\mu\nu}^I$ & $\mathcal{O}_{W\phi^4D^2}^{(2)}$   \\[0.2cm]
& $\text{i}\epsilon^{IJK} (\phi^{\dag} \sigma^I \phi) (D^{\mu} \phi^{\dag} \sigma^J D^{\nu} \phi) W_{\mu\nu}^K$  &
$\mathcal{O}_{W\phi^4D^2}^{(3)}$  & $\text{i}\epsilon^{IJK} (\phi^{\dag} \sigma^I \phi) (D^{\mu} \phi^{\dag} \sigma^J D^{\nu} \phi) \widetilde{W}_{\mu\nu}^K$ &  $\mathcal{O}_{W\phi^4D^2}^{(4)}$ \\[0.2cm]
& \textcolor{gray}{$(\phi^\dag \phi) D_\nu W^{I \mu\nu}(D_\mu\phi^\dagger \sigma^I \phi + \text{h.c.}) $ }& \textcolor{gray}{$\mathcal{O}_{W\phi^4 D^2}^{(5)}$}  &
 \textcolor{gray}{$(\phi^\dag \phi) D_\nu W^{I \mu\nu}(D_\mu\phi^\dagger \text{i} \sigma^I \phi + \text{h.c.})$} & \textcolor{gray}{$\mathcal{O}_{W\phi^4 D^2}^{(6)}$}  \\[0.2cm]
& \textcolor{gray}{$\epsilon^{IJK} (D_\mu \phi^\dag \sigma^I \phi) (\phi^\dag \sigma^J D_\nu \phi) W^{K \mu\nu} $} & \textcolor{gray}{$\mathcal{O}_{W\phi^4 D^2}^{(7)}$}  & $\text{i}(\phi^{\dag} \phi) (D^{\mu} \phi^{\dag} D^{\nu} \phi) B_{\mu\nu}$ & $\mathcal{O}_{B\phi^4D^2}^{(1)}$ \\[0.2cm]
&  $\text{i}(\phi^{\dag} \phi) (D^{\mu} \phi^{\dag} D^{\nu} \phi) \widetilde{B}_{\mu\nu}$ & $\mathcal{O}_{B\phi^4D^2}^{(2)}$ & \textcolor{gray}{$ (\phi^{\dag} \phi) D_{\nu} B^{\mu\nu} (D_\mu \phi^\dagger \text{i} \phi + \text{h.c.})$} & \textcolor{gray}{$\mathcal{O}_{B\phi^4D^2}^{(3)}$}\\[0.2cm]
\hline\\[-0.3cm]
  \rotatebox[origin=c]{90}{\textcolor{gray}{\boldmath{$\phi^2 D^6$}}}&\textcolor{gray}{ $D^2\phi^\dagger D_\mu D_\nu D^\mu D^\nu\phi$} & \textcolor{gray}{$\mathcal{O}_{\phi^2}$} & \\[0.4cm]
   \bottomrule
  \end{tabular}}
 \end{center}
  \caption{Green's basis of operators, part I. Operators in gray are redundant.}\label{tab:dim8ops1}
\end{table}

\begin{table}[h!]
 \begin{center}
  \resizebox{0.95\textwidth}{!}{
\begin{tabular}{cclcl}
   \toprule\\[-0.3cm]
   & \textbf{Operator} & \textbf{Notation} & \textbf{Operator} & \textbf{Notation}\\[0.5cm]

   \multirow{50}{*}{\rotatebox[origin=c]{90}{ {\boldmath{$X^2 \phi^2 D^2$}}}} &    {$(D^{\mu} \phi^{\dag} D^{\nu} \phi) W_{\mu\rho}^I W_{\nu}^{I \rho}$} &  {$\mathcal{O}_{W^2\phi^2D^2}^{(1)}$}  &
 {$(D^{\mu} \phi^{\dag} D_{\mu} \phi) W_{\nu\rho}^I W^{I \nu\rho}$} &  {$\mathcal{O}_{W^2\phi^2D^2}^{(2)}$} \\[0.2cm]
&   {$(D^{\mu} \phi^{\dag} D_{\mu} \phi) W_{\nu\rho}^I \widetilde{W}^{I \nu\rho}$} &  {$\mathcal{O}_{W^2\phi^2D^2}^{(3)}$} &
 {$i \epsilon^{IJK} (D^{\mu} \phi^{\dag} \sigma^I D^{\nu} \phi) W_{\mu\rho}^J W_{\nu}^{K \rho}$} &  {$\mathcal{O}_{W^2\phi^2D^2}^{(4)}$} \\[0.2cm]
&   {$\epsilon^{IJK} (D^{\mu} \phi^{\dag} \sigma^I D^{\nu} \phi) (W_{\mu\rho}^J \widetilde{W}_{\nu}^{K \rho} - \widetilde{W}_{\mu\rho}^J W_{\nu}^{K \rho})$} &  {$\mathcal{O}_{W^2\phi^2D^2}^{(5)}$} &   {$i \epsilon^{IJK} (D^{\mu} \phi^{\dag} \sigma^I D^{\nu} \phi) (W_{\mu\rho}^J \widetilde{W}_{\nu}^{K \rho} + \widetilde{W}_{\mu\rho}^J W_{\nu}^{K \rho})$} &  {$\mathcal{O}_{W^2\phi^2D^2}^{(6)}$} \\[0.2cm]
&\textcolor{gray}{$  i\epsilon^{IJK}(\phi^\dagger \sigma^I D^\nu\phi-D^\nu\phi^\dagger \sigma^I \phi) D_\mu W^{J\mu\rho}\widetilde{W}^{K}_{\nu\rho}$} & \textcolor{gray}{$ \mathcal{O}_{W^2\phi^2 D^2}^{(7)}  $} & \textcolor{gray}{$ \epsilon^{IJK} \phi^\dagger\sigma^I\phi D_\nu D_\mu W^{J\mu\rho}\widetilde{W}^{K\nu}_{\,\,\,\,\rho}$} & \textcolor{gray}{$ \mathcal{O}_{W^2\phi^2 D^2}^{(8)}$} \\[0.2cm]
& \textcolor{gray}{$i (\phi^\dagger D_\nu\phi-D_\nu\phi^\dagger \phi) D_\mu W^{I\mu\rho} \widetilde{W}^{I\nu}_{\,\,\,\,\rho}$} & \textcolor{gray}{$\mathcal{O}_{W^2\phi^2 D^2}^{(9)}$} & \textcolor{gray}{$(\phi^\dagger D_\nu\phi+D_\nu\phi^\dagger \phi) D_\mu W^{I\mu\rho} \widetilde{W}^{I\nu}_{\,\,\,\,\rho}$} & \textcolor{gray}{$  \mathcal{O}_{W^2\phi^2 D^2}^{(10)} $}  \\[0.2cm]
& \textcolor{gray}{$(\phi^\dagger D_\nu\phi+D_\nu\phi^\dagger \phi) D_\mu W^{I\mu\rho} W^{I\nu}_{\,\,\,\,\rho} $} & \textcolor{gray}{$ \mathcal{O}_{W^2\phi^2 D^2}^{(11)} $}  & \textcolor{gray}{$ i(\phi^\dagger D_\nu\phi-D_\nu\phi^\dagger \phi) D_\mu W^{I\mu\rho} W^{I\nu}_{\,\,\,\,\rho} $}& \textcolor{gray}{$\mathcal{O}_{W^2\phi^2 D^2}^{(12)}$} \\[0.2cm]
& \textcolor{gray}{$ \phi^\dagger\phi D_\mu W^{I\mu\rho} D_\nu W^{I\nu}_{\,\,\,\,\rho}$} & \textcolor{gray}{$  \mathcal{O}_{W^2\phi^2 D^2}^{(13)}$} & \textcolor{gray}{$(D_\mu\phi^\dagger\phi+\phi^\dagger D_\mu\phi) W^{I\nu\rho} D^\mu W^{I}_{\nu\rho} $} & \textcolor{gray}{$ \mathcal{O}_{W^2\phi^2 D^2}^{(14)}$} \\[0.2cm]
& \textcolor{gray}{$i(D_\mu\phi^\dagger\phi-\phi^\dagger D_\mu\phi) W^{I\nu\rho} D^\mu W^{I}_{\nu\rho} $} & \textcolor{gray}{$ \mathcal{O}_{W^2\phi^2 D^2}^{(15)}$} & \textcolor{gray}{$(D_\mu\phi^\dagger\phi+\phi^\dagger D_\mu\phi) D^\mu W^{I\nu\rho}\widetilde{W}^{I}_{\nu\rho} $} & \textcolor{gray}{$ \mathcal{O}_{W^2\phi^2 D^2}^{(16)}  $} \\[0.2cm]
& \textcolor{gray}{$ i(D_\mu\phi^\dagger\phi-\phi^\dagger D_\mu\phi) D^\mu W^{I\nu\rho}\widetilde{W}^{I}_{\nu\rho} $} & \textcolor{gray}{$ \mathcal{O}_{W^2\phi^2 D^2}^{(17)}$} & \textcolor{gray}{$ \epsilon^{IJK}(\phi^\dagger \sigma^I D^\nu\phi+D^\nu\phi^\dagger \sigma^I \phi) D_\mu W^{J\mu\rho}W^{K}_{\nu\rho} $} & \textcolor{gray}{$ \mathcal{O}_{W^2\phi^2 D^2}^{(18)}  $} \\[0.2cm]
& \textcolor{gray}{$i\epsilon^{IJK}(\phi^\dagger \sigma^I D^\nu\phi-D^\nu\phi^\dagger \sigma^I \phi) D_\mu W^{J\mu\rho}W^{K}_{\nu\rho}$} & \textcolor{gray}{$ \mathcal{O}_{W^2\phi^2 D^2}^{(19)}$} & & \\[0.2cm]
& \\[0.05cm]
&   {$(D^{\mu} \phi^{\dag} \sigma^I D_{\mu} \phi) B_{\nu\rho} W^{I \nu\rho}$} &  {$\mathcal{O}_{WB\phi^2D^2}^{(1)}$} &   {$(D^{\mu} \phi^{\dag} \sigma^I D_{\mu} \phi) B_{\nu\rho} \widetilde{W}^{I \nu\rho}$} &  {$\mathcal{O}_{WB\phi^2D^2}^{(2)}$} \\[0.2cm]
&   {$i (D^{\mu} \phi^{\dag} \sigma^I D^{\nu} \phi) (B_{\mu\rho} W_{\nu}^{I \rho} - B_{\nu\rho} W_{\mu}^{I\rho})$} &  {$\mathcal{O}_{WB\phi^2D^2}^{(3)}$}  &   {$(D^{\mu} \phi^{\dag} \sigma^I D^{\nu} \phi) (B_{\mu\rho} W_{\nu}^{I \rho} + B_{\nu\rho} W_{\mu}^{I\rho})$} & 
 {$\mathcal{O}_{WB\phi^2D^2}^{(4)}$} \\[0.2cm]
&  {$i (D^{\mu} \phi^{\dag} \sigma^I D^{\nu} \phi) (B_{\mu\rho} \widetilde{W}_\nu^{^I \rho} - B_{\nu\rho} \widetilde{W}_\mu^{^I \rho})$} &  {$\mathcal{O}_{WB\phi^2D^2}^{(5)}$}   &  {$(D^{\mu} \phi^{\dag} \sigma^I D^{\nu} \phi) (B_{\mu\rho} \widetilde{W}_\nu^{^I \rho} + B_{\nu\rho} \widetilde{W}_\mu^{^I \rho})$} &  {$\mathcal{O}_{WB\phi^2D^2}^{(6)}$} \\[0.2cm]
&  \textcolor{gray}{$i(\phi^\dagger\sigma^I D^\mu\phi-D^\mu\phi^\dagger\sigma^I\phi) D_\mu B^{\nu\rho} W^{I}_{\nu\rho}$} &  \textcolor{gray}{$\mathcal{O}_{WB\phi^2D^2}^{(7)}$}   &  \textcolor{gray}{$ (\phi^\dagger\sigma^I D^\nu\phi+D^\nu\phi^\dagger\sigma^I\phi) D_\mu B^{\mu\rho} W^{I}_{\nu\rho}$} &  \textcolor{gray}{$\mathcal{O}_{WB\phi^2D^2}^{(8)}$} \\[0.2cm]
&  \textcolor{gray}{$i(\phi^\dagger\sigma^I D^\nu\phi-D^\nu\phi^\dagger\sigma^I\phi) D_\mu B^{\mu\rho} W^{I}_{\nu\rho}$} &  \textcolor{gray}{$\mathcal{O}_{WB\phi^2D^2}^{(9)}$}   & \textcolor{gray} {$(\phi^\dagger\sigma^I\phi) D^\mu B_{\mu\rho} D_\nu W^{I\nu\rho}$} &  \textcolor{gray}{$\mathcal{O}_{WB\phi^2D^2}^{(10)}$} \\[0.2cm]
&  \textcolor{gray}{$(D_\nu\phi^\dagger\sigma^I\phi+\phi^\dagger\sigma^I D_\nu\phi) B_{\mu\rho} D^\mu W^{I\nu\rho}$} &  \textcolor{gray}{$\mathcal{O}_{WB\phi^2D^2}^{(11)}$}   & \textcolor{gray} {$ i(D_\nu\phi^\dagger\sigma^I\phi-\phi^\dagger\sigma^I D_\nu\phi) B_{\mu\rho} D^\mu W^{I\nu\rho}$} &  \textcolor{gray}{$\mathcal{O}_{WB\phi^2D^2}^{(12)}$} \\[0.2cm]
&  \textcolor{gray}{$(\phi^\dagger\sigma^I\phi) B_{\mu\rho}  D_\nu D^\mu W^{I\nu\rho}$} &  \textcolor{gray}{$\mathcal{O}_{WB\phi^2D^2}^{(13)}$}   &  \textcolor{gray}{$i(D_\nu\phi^\dagger\sigma^I\phi-\phi^\dagger\sigma^I D_\nu\phi) D^\mu B_{\mu\rho} \widetilde{W}^{I\nu\rho} $} &  \textcolor{gray}{$\mathcal{O}_{WB\phi^2D^2}^{(14)}$} \\[0.2cm]
&  \textcolor{gray}{$ i (\phi^\dagger\sigma^I D_\mu\phi-D_\mu\phi^\dagger\sigma^I\phi) D^\mu B_{\nu\rho} \widetilde{W}^{I\nu\rho}$} & \textcolor{gray} {$\mathcal{O}_{WB\phi^2D^2}^{(15)}$}   &  \textcolor{gray}{$(\phi^\dagger\sigma^I\phi) (D^2 B^{\nu\rho}) \widetilde{W}^{I}_{\nu\rho} $} &  \textcolor{gray}{$\mathcal{O}_{WB\phi^2D^2}^{(16)}$} \\[0.2cm]
&  \textcolor{gray}{$ (\phi^\dagger\sigma^I\phi) (D^\rho D_\mu W^{I\mu\nu}) \widetilde{B}_{\nu\rho}$} &  \textcolor{gray}{$\mathcal{O}_{WB\phi^2D^2}^{(17)}$}  &  \textcolor{gray}{$i(D^\nu\phi^\dagger\sigma^I\phi-\phi^\dagger\sigma^I D^\nu\phi) \widetilde{B}^{\mu\rho} D_\mu W^I_{\nu\rho}$} &  \textcolor{gray}{$\mathcal{O}_{WB\phi^2D^2}^{(18)}$} \\[0.2cm]
& \textcolor{gray}{$(D^\nu\phi^\dagger\sigma^I\phi+\phi^\dagger\sigma^I D^\nu\phi) \widetilde{B}^{\mu\rho} D_\mu W^I_{\nu\rho}$} &  \textcolor{gray}{$\mathcal{O}_{WB\phi^2D^2}^{(19)}$}  \\[0.2cm]
& \\[0.05cm]
&  {$(D^{\mu} \phi^{\dag} D^{\nu} \phi) B_{\mu\rho} B_{\nu}^{\,\,\,\rho}$} &  {$\mathcal{O}_{B^2\phi^2D^2}^{(1)}$} &
  {$(D^{\mu} \phi^{\dag} D_{\mu} \phi) B_{\nu\rho} B^{\nu\rho}$} &  {$\mathcal{O}_{B^2\phi^2D^2}^{(2)}$}   \\[0.2cm]
&  {$(D^{\mu} \phi^{\dag} D_{\mu} \phi) B_{\nu\rho} \widetilde{B}^{\nu\rho}$} &  {$\mathcal{O}_{B^2\phi^2D^2}^{(3)}$} & \textcolor{gray}{$(D_\mu\phi^\dagger \phi+\phi^\dagger D_\mu\phi) D_\nu B^{\mu\rho} B^\nu_{\,\,\rho} $} & \textcolor{gray}{$\mathcal{O}_{B^2\phi^2 D^2}^{(4)}  $}  \\[0.2cm] 
&  \textcolor{gray}{$ i (\phi^\dagger D_\mu D_\nu\phi-D_\mu D_\nu \phi^\dagger \phi) B^{\mu\rho} B^{\nu}_{\,\,\rho}$} &  \textcolor{gray}{$\mathcal{O}_{B^2\phi^2D^2}^{(5)}$} & \textcolor{gray}{$\phi^\dagger\phi D_\mu D_\nu B^{\mu\rho} B^{\nu}_{\,\,\rho} $} & \textcolor{gray}{$\mathcal{O}_{B^2\phi^2 D^2}^{(6)}  $}  \\[0.2cm] 
&  \textcolor{gray}{$ i (\phi^\dagger D_\nu\phi-D_\nu\phi^\dagger\phi) D_\mu B^{\mu\rho} B^{\nu}_{\,\,\rho}$} &  \textcolor{gray}{$\mathcal{O}_{B^2\phi^2D^2}^{(7)}$} & \textcolor{gray}{$(\phi^\dagger D_\nu\phi+D_\nu\phi^\dagger\phi) D_\mu B^{\mu\rho} B^\nu_{\,\,\rho}$} & \textcolor{gray}{$\mathcal{O}_{B^2\phi^2 D^2}^{(8)}  $}  \\[0.2cm] 
&  \textcolor{gray}{$ (\phi^\dagger D^2\phi+D^2\phi^\dagger\phi) B^{\nu\rho} \widetilde{B}_{\nu\rho}$} &  \textcolor{gray}{$\mathcal{O}_{B^2\phi^2D^2}^{(9)}$} & \textcolor{gray}{$ i (\phi^\dagger D^2\phi-D^2\phi^\dagger\phi) B^{\nu\rho} \widetilde{B}_{\nu\rho}$} & \textcolor{gray}{$\mathcal{O}_{B^2\phi^2 D^2}^{(10)}  $}  \\[0.2cm] 
&  \textcolor{gray}{$ (\phi^\dagger D_\nu\phi+D_\nu\phi^\dagger\phi) D_\mu B^{\mu\rho} \widetilde{B}^{\nu}_{\,\,\rho}$} &  \textcolor{gray}{$\mathcal{O}_{B^2\phi^2 D^2}^{(11)}$} & \textcolor{gray}{$  i(\phi^\dagger D_\nu\phi - D_\nu\phi^\dagger\phi) D_\mu B^{\mu\rho} \widetilde{B}^{\nu}_{\,\,\rho}$} & \textcolor{gray}{$\mathcal{O}_{B^2\phi^2 D^2}^{(12)}  $}  \\[0.2cm] 
& \\[0.05cm]
&  {$(D^\mu\phi^\dagger D_\nu\phi) G^A_{\mu\rho} G^{A\nu\rho}$} &  {$\mathcal{O}_{G^2\phi^2D^2}^{(1)}$} &
  {$(D^\mu\phi^\dagger D_\mu\phi) G^A_{\nu\rho} G^{A\nu\rho}$} &  {$\mathcal{O}_{G^2\phi^2D^2}^{(2)}$}   \\[0.2cm]
&  {$ (D^\mu\phi^\dagger D_\mu\phi) G^A_{\nu\rho} \widetilde{G}^{A\nu\rho}$} &  {$\mathcal{O}_{G^2\phi^2D^2}^{(3)}$} & \textcolor{gray}{$(D_\mu\phi^\dagger \phi+\phi^\dagger D_\mu\phi) D_\nu G^{A\mu\rho} G^{A\nu}_{\,\,\rho}$} & \textcolor{gray}{$\mathcal{O}_{G^2\phi^2 D^2}^{(4)}  $}  \\[0.2cm] 
&  \textcolor{gray}{$ i (\phi^\dagger D_\mu D_\nu\phi-D_\mu D_\nu \phi^\dagger \phi) G^{A\mu\rho} G^{A\nu}_{\,\,\rho}$} &  \textcolor{gray}{$\mathcal{O}_{G^2\phi^2D^2}^{(5)}$} & \textcolor{gray}{$\phi^\dagger\phi D_\mu D_\nu G^{A\mu\rho} G^{A\nu}_{\,\,\rho}$} & \textcolor{gray}{$\mathcal{O}_{G^2\phi^2 D^2}^{(6)}  $}  \\[0.2cm] 
&  \textcolor{gray}{$ i (\phi^\dagger D_\nu\phi-D_\nu\phi^\dagger\phi) D_\mu G^{A\mu\rho} G^{A\nu}_{\,\,\rho}$} &  \textcolor{gray}{$\mathcal{O}_{G^2\phi^2D^2}^{(7)}$} & \textcolor{gray}{$ (\phi^\dagger D_\nu\phi+D_\nu\phi^\dagger\phi) D_\mu G^{A\mu\rho} G^{A\nu}_{\,\,\rho}$} & \textcolor{gray}{$\mathcal{O}_{G^2\phi^2 D^2}^{(8)}  $}  \\[0.2cm] 
&  \textcolor{gray}{$ (\phi^\dagger D^2\phi+D^2\phi^\dagger\phi) G^{A\nu\rho} \widetilde{G}^{A\nu}_{\,\,\rho}$} &  \textcolor{gray}{$\mathcal{O}_{G^2\phi^2D^2}^{(9)}$} & \textcolor{gray}{$ i (\phi^\dagger D^2\phi-D^2\phi^\dagger\phi) G^{A\nu\rho} \widetilde{G}^A_{\nu\rho}$} & \textcolor{gray}{$\mathcal{O}_{G^2\phi^2 D^2}^{(10)}  $}  \\[0.2cm] 
&  \textcolor{gray}{$ (\phi^\dagger D_\nu\phi+D_\nu\phi^\dagger\phi) D_\mu G^{A\mu\rho} \widetilde{G}^{A\nu}_{\,\,\rho}$} &  \textcolor{gray}{$\mathcal{O}_{G^2\phi^2 D^2}^{(11)}$} & \textcolor{gray}{$  i(\phi^\dagger D_\nu\phi - D_\nu\phi^\dagger\phi) D_\mu G^{A\mu\rho} \widetilde{G}^{A\nu}_{\,\,\rho}$} & \textcolor{gray}{$\mathcal{O}_{G^2\phi^2 D^2}^{(12)}  $}  \\[0.2cm] 

   \bottomrule
  \end{tabular}}
 \end{center}
 \caption{Green's basis of operators, part II.}\label{tab:dim8ops2}
\end{table}
\begin{table}[h!]
 \begin{center}
  \resizebox{0.8\textwidth}{!}{
\begin{tabular}{cclcl}
   \toprule\\[-0.3cm]
   & \textbf{Operator} & \textbf{Notation} & \textbf{Operator} & \textbf{Notation}\\[0.5cm]

   \multirow{18}{*}{\rotatebox[origin=c]{90}{ {\boldmath{$X^4,\, X^3 X^{\prime}$}}}} &    { $(G_{\mu\nu}^A G^{A\mu\nu}) (G_{\rho\sigma}^B G^{B\rho\sigma})$} &  {$Q_{G^4}^{(1)}$}  &
 {$(G_{\mu\nu}^A \widetilde{G}^{A\mu\nu}) (G_{\rho\sigma}^B \widetilde{G}^{B\rho\sigma})$} &  {$Q_{G^4}^{(2)}$} \\[0.2cm]
&   {$(G_{\mu\nu}^A G^{B\mu\nu}) (G_{\rho\sigma}^A G^{B\rho\sigma})$} &  {$Q_{G^4}^{(3)}$ } &
 {$(G_{\mu\nu}^A \widetilde{G}^{B\mu\nu}) (G_{\rho\sigma}^A \widetilde{G}^{B\rho\sigma})$} &  {$Q_{G^4}^{(4)}$} \\[0.2cm]
&   {$(G_{\mu\nu}^A G^{A\mu\nu}) (G_{\rho\sigma}^B \widetilde{G}^{B\rho\sigma})$} &  {$Q_{G^4}^{(5)}$} &   {$(G_{\mu\nu}^A G^{B\mu\nu}) (G_{\rho\sigma}^A \widetilde{G}^{B\rho\sigma})$} &  {$Q_{G^4}^{(6)}$} \\[0.2cm]
& $d^{ABE} d^{CDE} (G_{\mu\nu}^A G^{B\mu\nu}) (G_{\rho\sigma}^C G^{D\rho\sigma})$ & $Q_{G^4}^{(7)}$ & $ d^{ABE} d^{CDE} (G_{\mu\nu}^A \widetilde{G}^{B\mu\nu}) (G_{\rho\sigma}^C \widetilde{G}^{D\rho\sigma})$ & $Q_{G^4}^{(8)}$  \\[0.2cm]
& $d^{ABE} d^{CDE} (G_{\mu\nu}^A G^{B\mu\nu}) (G_{\rho\sigma}^C \widetilde{G}^{D\rho\sigma})$ & $ Q_{G^4}^{(9)} $  \\
&\\[0.05cm]
& $(W_{\mu\nu}^I W^{I\mu\nu}) (W_{\rho\sigma}^J W^{J\rho\sigma})$ & $  Q_{W^4}^{(1)}$ & $(W_{\mu\nu}^I \widetilde{W}^{I\mu\nu}) (W_{\rho\sigma}^J \widetilde{W}^{J\rho\sigma})$ & $Q_{W^4}^{(2)}$ \\[0.2cm]
& $(W_{\mu\nu}^I W^{J\mu\nu}) (W_{\rho\sigma}^I W^{J\rho\sigma}) $ & $ Q_{W^4}^{(3)}$ & $(W_{\mu\nu}^I \widetilde{W}^{J\mu\nu}) (W_{\rho\sigma}^I \widetilde{W}^{J\rho\sigma}) $ & $ Q_{W^4}^{(4)} $ \\[0.2cm]
& $(W_{\mu\nu}^I W^{I\mu\nu}) (W_{\rho\sigma}^J \widetilde{W}^{J\rho\sigma}) $ & $Q_{W^4}^{(5)}$ & $(W_{\mu\nu}^I W^{J\mu\nu}) (W_{\rho\sigma}^I \widetilde{W}^{J\rho\sigma})$ & $ Q_{W^4}^{(6)} $ \\
& \\[0.05cm]
&   {$(B_{\mu\nu} B^{\mu\nu}) (B_{\rho\sigma} B^{\rho\sigma})$} &  {$Q_{B^4}^{(1)}$} &   {$(B_{\mu\nu} \widetilde{B}^{\mu\nu}) (B_{\rho\sigma} \widetilde{B}^{\rho\sigma})$} &  {$Q_{B^4}^{(2)}$} \\[0.2cm]
&   {$(B_{\mu\nu} B^{\mu\nu}) (B_{\rho\sigma} \widetilde{B}^{\rho\sigma})$} &  {$Q_{B^4}^{(3)}$}  \\
&\\[0.05cm]
&  {$d^{ABC} (B_{\mu\nu} G^{A\mu\nu}) (G_{\rho\sigma}^B G^{C\rho\sigma})$} &  {$Q_{G^3B}^{(1)}$}   &  {$d^{ABC} (B_{\mu\nu} \widetilde{G}^{A\mu\nu}) (G_{\rho\sigma}^B \widetilde{G}^{C\rho\sigma})$} &  {$Q_{G^3B}^{(2)}$} \\[0.2cm]
&  {$d^{ABC} (B_{\mu\nu} \widetilde{G}^{A\mu\nu}) (G_{\rho\sigma}^B G^{C\rho\sigma})$} &  {$Q_{G^3B}^{(3)}$}   &  {$d^{ABC} (B_{\mu\nu} G^{A\mu\nu}) (G_{\rho\sigma}^B \widetilde{G}^{C\rho\sigma})$} &  {$Q_{G^3B}^{(4)}$} \\[0.2cm]
 \hline\\[-0.3cm]
   \multirow{18}{*}{\rotatebox[origin=c]{90}{ {\boldmath{$X^2 X^{\prime 2}$}}}} &    {$(W_{\mu\nu}^I W^{I\mu\nu}) (G_{\rho\sigma}^A G^{A\rho\sigma})$} &  {$Q_{G^2W^2}^{(1)}$}  &
 {$(W_{\mu\nu}^I \widetilde{W}^{I\mu\nu}) (G_{\rho\sigma}^A \widetilde{G}^{A\rho\sigma})$} &  {$Q_{G^2W^2}^{(2)}$} \\[0.2cm]
&    {$(W_{\mu\nu}^I G^{A\mu\nu}) (W_{\rho\sigma}^I G^{A\rho\sigma})$} &  {$Q_{G^2W^2}^{(3)}$}  &
 {$(W_{\mu\nu}^I \widetilde{G}^{A\mu\nu}) (W_{\rho\sigma}^I \widetilde{G}^{A\rho\sigma})$} &  {$Q_{G^2W^2}^{(4)}$} \\[0.2cm]
&    {$(W_{\mu\nu}^I \widetilde{W}^{I\mu\nu}) (G_{\rho\sigma}^A G^{A\rho\sigma})$} &  {$Q_{G^2W^2}^{(5)}$}  &
 {$(W_{\mu\nu}^I W^{I\mu\nu}) (G_{\rho\sigma}^A \widetilde{G}^{A\rho\sigma})$} &  {$Q_{G^2W^2}^{(6)}$} \\[0.2cm]
&    {$(W_{\mu\nu}^I G^{A\mu\nu}) (W_{\rho\sigma}^I \widetilde{G}^{A\rho\sigma})$} &  {$Q_{G^2W^2}^{(7)}$}  \\
\\[0.05cm]
&    {$(B_{\mu\nu} B^{\mu\nu}) (G_{\rho\sigma}^A G^{A\rho\sigma})$} &  {$Q_{G^2B^2}^{(1)}$}  &
 {$(B_{\mu\nu} \widetilde{B}^{\mu\nu}) (G_{\rho\sigma}^A \widetilde{G}^{A\rho\sigma})$} &  {$Q_{G^2B^2}^{(2)}$} \\[0.2cm]
&    {$(B_{\mu\nu} G^{A\mu\nu}) (B_{\rho\sigma} G^{A\rho\sigma})$} &  {$Q_{G^2B^2}^{(3)}$}  &
 {$(B_{\mu\nu} \widetilde{G}^{A\mu\nu}) (B_{\rho\sigma} \widetilde{G}^{A\rho\sigma})$} &  {$Q_{G^2B^2}^{(4)}$} \\[0.2cm]
&    {$(B_{\mu\nu} \widetilde{B}^{\mu\nu}) (G_{\rho\sigma}^A G^{A\rho\sigma})$} &  {$Q_{G^2B^2}^{(5)}$}  &
 {$(B_{\mu\nu} B^{\mu\nu}) (G_{\rho\sigma}^A \widetilde{G}^{A\rho\sigma})$} &  {$Q_{G^2B^2}^{(6)}$} \\[0.2cm]
&    {$(B_{\mu\nu} G^{A\mu\nu}) (B_{\rho\sigma} \widetilde{G}^{A\rho\sigma})$} &  {$Q_{G^2B^2}^{(7)}$} \\
\\[0.05cm]
&    {$(B_{\mu\nu} B^{\mu\nu}) (W_{\rho\sigma}^I W^{I\rho\sigma})$} &  {$Q_{W^2B^2}^{(1)}$}  &
 {$(B_{\mu\nu} \widetilde{B}^{\mu\nu}) (W_{\rho\sigma}^I \widetilde{W}^{I\rho\sigma})$} &  {$Q_{W^2B^2}^{(2)}$} \\[0.2cm]
&    {$(B_{\mu\nu} W^{I\mu\nu}) (B_{\rho\sigma} W^{I\rho\sigma})$} &  {$Q_{W^2B^2}^{(3)}$}  &
 {$(B_{\mu\nu} \widetilde{W}^{I\mu\nu}) (B_{\rho\sigma} \widetilde{W}^{I\rho\sigma})$} &  {$Q_{W^2B^2}^{(4)}$} \\[0.2cm]
&    {$(B_{\mu\nu} \widetilde{B}^{\mu\nu}) (W_{\rho\sigma}^I W^{I\rho\sigma})$} &  {$Q_{W^2B^2}^{(5)}$}  &
 {$(B_{\mu\nu} B^{\mu\nu}) (W_{\rho\sigma}^I \widetilde{W}^{I\rho\sigma})$} &  {$Q_{W^2B^2}^{(6)}$} \\[0.2cm]
&    {$(B_{\mu\nu} W^{I\mu\nu}) (B_{\rho\sigma} \widetilde{W}^{I\rho\sigma})$} &  {$Q_{W^2B^2}^{(7)}$}  \\[0.2cm]

   \bottomrule
  \end{tabular}}
 \end{center}
 \caption{Green's basis of operators, part III.}\label{tab:dim8ops3}
\end{table}
\begin{table}[h!]
 \begin{center}
  \resizebox{0.8\textwidth}{!}{
\begin{tabular}{cclcl}
   \toprule\\[-0.3cm]
   & \textbf{Operator} & \textbf{Notation} & \textbf{Operator} & \textbf{Notation}\\[0.5cm]

   \multirow{12}{*}{\rotatebox[origin=c]{90}{ {\boldmath{$X^3 D^2$}}}} &    \textcolor{gray}{$B_{\mu\nu} D_\rho W^{I\mu\nu} D_\sigma W^{I\rho\sigma}$} &  \textcolor{gray}{$ \mathcal{O}_{W^2B D^2}^{(1)}$}  &
 \textcolor{gray}{$ B_{\mu\nu} (D^2 W^{I\mu\rho}) W^{I\nu}_{\,\,\,\,\,\,\rho}$} &  \textcolor{gray}{$\mathcal{O}_{W^2B D^2}^{(2)}$} \\[0.2cm]
&    \textcolor{gray}{$ \widetilde{B}_{\mu\nu} D_\rho W^{I\mu\nu} D_\sigma W^{I\rho\sigma}$} &  \textcolor{gray}{$ \mathcal{O}_{W^2B D^2}^{(3)}$}  &
 \textcolor{gray}{$\widetilde{B}_{\mu\nu} (D^2 W^{I\mu\rho}) W^{I\nu}_{\,\,\,\,\,\,\rho}$} & \textcolor{gray} {$\mathcal{O}_{W^2B D^2}^{(4)} $} \\
&\\[0.05cm]
&    \textcolor{gray}{$B_{\mu\nu} D_\rho G^{A\mu\nu} D_\sigma G^{A\rho\sigma}$} &  \textcolor{gray}{$ \mathcal{O}_{G^2B D^2}^{(1)}$}  &
\textcolor{gray} {$ B_{\mu\nu} (D^2 G^{A\mu\rho}) G^{A\nu}_{\,\,\,\,\,\,\rho}$} & \textcolor{gray} {$\mathcal{O}_{G^2B D^2}^{(2)}$} \\[0.2cm]
&   \textcolor{gray} {$ \widetilde{B}_{\mu\nu} D_\rho G^{A\mu\nu} D_\sigma G^{A\rho\sigma}$} &  \textcolor{gray}{$ \mathcal{O}_{G^2B D^2}^{(3)}$}  &
 \textcolor{gray}{$\widetilde{B}_{\mu\nu} (D^2 G^{A\mu\rho}) G^{A\nu}_{\,\,\,\,\,\,\rho}$} &  \textcolor{gray}{$\mathcal{O}_{G^2B D^2}^{(4)} $} \\
&\\[0.05cm]
&    \textcolor{gray}{$ \epsilon^{IJK} W^I_{\mu\nu} D_\rho W^{J\mu\nu} D_\sigma W^{K\rho\sigma}$} & \textcolor{gray} {$ \mathcal{O}_{W^3 D^2}^{(1)} $}  &
 \textcolor{gray}{$ \epsilon^{IJK} W^I_{\mu\nu} D_\rho W^{J\rho\mu} D_\sigma W^{K\sigma\nu}$} &  \textcolor{gray}{$\mathcal{O}_{W^3 D^2}^{(2)}$} \\[0.2cm]
&   \textcolor{gray} {$\epsilon^{IJK}\widetilde{W}^I_{\mu\nu} D_\rho W^{J\mu\nu} D_\sigma W^{K\rho\sigma}$} &  \textcolor{gray}{$ \mathcal{O}_{W^3 D^2}^{(3)} $}  &
 \textcolor{gray}{$\epsilon^{IJK} \widetilde{W}^I_{\mu\nu} D_\rho W^{J\rho\mu} D_\sigma W^{K\sigma\nu}$} &  \textcolor{gray}{$\mathcal{O}_{W^3 D^2}^{(4)} $} \\
&\\[0.05cm]
&    \textcolor{gray}{$ f^{ABC} G^A_{\mu\nu} D_\rho G^{B\mu\nu} D_\sigma G^{C\rho\sigma}$} & \textcolor{gray} {$ \mathcal{O}_{G^3 D^2}^{(1)} $}  &
 \textcolor{gray}{$ f^{ABC} G^A_{\mu\nu} D_\rho G^{B\rho\mu} D_\sigma G^{C\sigma\nu} $} &  \textcolor{gray}{$\mathcal{O}_{G^3 D^2}^{(2)}$} \\[0.2cm]
&  \textcolor{gray}  {$f^{ABC}\widetilde{G}^A_{\mu\nu} D_\rho G^{B\mu\nu} D_\sigma G^{C\rho\sigma}$} & \textcolor{gray} {$ \mathcal{O}_{G^3 D^2}^{(3)} $}  &
 \textcolor{gray}{$f^{ABC} \widetilde{G}^A_{\mu\nu} D_\rho G^{B\rho\mu} D_\sigma G^{C\sigma\nu} $} &  \textcolor{gray}{$\mathcal{O}_{G^3 D^2}^{(4)} $} \\[0.2cm]

 \hline\\[-0.3cm]
  \multirow{1}{*}{\rotatebox[origin=c]{90}{ {\boldmath{$X^2 D^4$}}}} &    \textcolor{gray}{$(D_\sigma D_\mu B^{\mu\nu}) (D^\sigma D^\rho B_{\rho\nu})$} &  \textcolor{gray}{$  \mathcal{O}_{B^2 D^4} $}  &
 \textcolor{gray}{$ (D_\sigma D_\mu W^{I\mu\nu}) (D^\sigma D^\rho W^I_{\rho\nu})$} &  \textcolor{gray}{$ \mathcal{O}_{W^2 D^4} $} \\[0.2cm]
  &    \textcolor{gray}{$ (D_\sigma D_\mu G^{A\mu\nu}) (D^\sigma D^\rho G^A_{\rho\nu})$} &  \textcolor{gray}{$  \mathcal{O}_{G^2 D^4} $} \\[0.2cm]

 \bottomrule
  \end{tabular}}
 \end{center}
 \caption{Green's basis of operators, part IV.}\label{tab:dim8ops4}
\end{table}

\clearpage
\bibliographystyle{style} 
\bibliography{refs} 

\end{document}